\documentclass[a4paper,11pt]{article}
\pdfoutput=1 

\usepackage{jheppub} 
\usepackage{graphicx,color}
\numberwithin{equation}{section}

 \usepackage{hyperref}
\makeatletter
\newcommand{\thickhline}{%
\noalign {\ifnum 0=`}\fi \hrule height 1pt
\futurelet \reserved@a \@xhline}
\RequirePackage[normalem]{ulem} 
\RequirePackage{color}\definecolor{RED}{rgb}{1,0,0}\definecolor{BLUE}{rgb}{0,0,1} 

\begin{document}

\title{\bf \Large $R_{K^{(*)}}$ anomaly in  type-III  2HDM  }

\author[a]{A.~Arhrib}
\author[b,c,d]{R.~Benbrik}
\author[e]{C.~H.~Chen}
\author[d]{J.~K.~Parry}
\author[f]{L.~Rahili}
\author[g]{S.~Semlali}
\author[c,d]{Q.~S.~Yan}

\affiliation[a]{Abdelmalek Essaadi University, Faculty of Sciences and techniques,Tanger, Morocco}
\affiliation[b]{MSISM Team, Facult\'e Polydisciplinaire de Safi, 
Sidi Bouzid, Morocco}
\affiliation[c]{School of Physics Sciences, University of Chinese, Chinese Academy of Sciences, Beijing 100039, P.R China}
\affiliation[d]{Center for future high energy physics, 
Chinese Academy of Sciences, Beijing 100039, P.R China}
\affiliation[e]{Department of Physics, National Cheng-Kung University, Tainan 70101, Taiwan}
\affiliation[f]{LMTI, Faculty of Sciences, Agadir University, B.P 8106, Morocco.}
\affiliation[g]{LPHEA, Faculty of Science Semlalia, Marrakech, 430079, Morocco.}
\affiliation[c]{School of Physics Sciences, University of Chinese, 
Chinese Academy of Sciences, Beijing 100039, P.R China}

\emailAdd{aarhrib@gmail.com}
\emailAdd{r.benbrik@uca.ac.ma}
\emailAdd{physchen@mail.ncku.edu.tw}
\emailAdd{jkparry@tsinghua.edu.cn}
\emailAdd{rahililarbi@gmail.com}
\emailAdd{s.seemlali@gmail.com}
\emailAdd{yanqishu@ucas.ac.cn}

\date{}

\abstract{
Recent experimental results provided by the CMS and LHCb, Belle and BaBar  
collaborations are showing a tension with the SM predictions in $R_{K^{(*)}}$, which might call for an explanation from new physics. In this work, we examine this tension in the type-III two-Higgs doublet models. We focus on the contributions of charged Higgs boson to the observable(s) $R_{K^{(*)}}$  and other rare processes 
$\Delta M_q$ ($q=s,d$), $B \to  X_s \gamma$  $B_s \to \mu^+ \mu^{-}$ and $B_q
\to X_s \mu^+ \mu^{-}$, which are governed by the same effective
Hamiltonian. It is found that regions of large $\tan\beta$ and light charged
Higgs mass $m_{H^\pm}$ can explain the measured value of $R_{K^{(*)}}$ 
 and accommodate other B physics data as well. In contrast, 
the type-II two-Higgs doublet model can not.}
\keywords{: LHCb, Belle, type-III 2HDM, $R_{K^{(*)}}$.}
\maketitle

\section{Introduction}
The Standard model (SM) has been completed by the discovery of the last
missing piece, the Higgs boson, at the Large Hadron Collider (LHC) at
CERN~\cite{Aad:2012tfa,Chatrchyan:2012ufa}. Up to now, significant direct
evidence for new physics beyond the SM need to be found at LHC with high
luminosity option. Nevertheless, CMS and LHCb collaborations have presented the analysis for the rare processes like $B\to K^* \mu^+\mu^-$, $B_s \to \mu^+\mu^-$ and $R_{K^{(*)}} = BR(B\to K^{(*)}\mu^+\mu^-)/BR(B\to K^{(*)}e^+e^-)$ based on the full set Run-1 data sets. Such precision measurements can serve as a guideline in the exploration of possible new physics.

A deviations from the SM predictions \cite{Egede:2008uy} reported by LHCb~\cite{Aaij:2013qta,LHCb:2015dla} and CMS~\cite{Khachatryan:2015isa}, later confirmed by Belle~\cite{Wehle:2016yoi} has shown in the rare process $B\to K^* \mu^+\mu^-$, mainly in an angular observable called $P_5^\prime$~\cite{Descotes-Genon:2013vna} with a significance of $2$--$3\,\sigma$ depending on the assumptions of hadronic uncertainties~
\cite{Descotes-Genon:2014uoa,Altmannshofer:2014rta,Jager:2014rwa}. Also a $3.5 \sigma$ discrepancy in the decay $B_s\to\phi\mu^+\mu^-$~\cite{Aaij:2013aln} has been reported by the LHCb collaboration, where the SM prediction are based on lattice QCD computation~\cite{Horgan:2013pva,Horgan:2015vla} and the light-cone sum rules~\cite{Straub:2015ica}. Furthermore, a violation of lepton flavour universality has been observed by the LHCb collaborations~
\cite{Aaij:2014ora}, to be precise, $ R_K=0.745^{+0.090}_{-0.074}\pm 0.036\,,$ in the range $1\,{\rm GeV^2}<q^2<6\,{\rm GeV^2}$, and $ R^{low}_{K^*}=0.660^{+0.110}_{-0.070}\pm 0.024\,,$ in the range $0.045\,{\rm GeV^2}<q^2<1.1\,{\rm GeV^2}$ which deviates by $2.6\,\sigma$ and $2.1\,\sigma$ from the SM precision prediction $R^{\rm SM}_{K^{(*)}}=1.0003(0.99) \pm 0.0001$, respectively~\cite{Bobeth:2007dw}. When these anomalies are combined with  other observables for the rare processes $b\to s \mu^+\mu^-$ transitions, it is found that a scenario with NP in $C_9^{\mu}$ (but not in $C_9^{e}$) is preferred. The best fit yielded a central value $C_9^{\mu} \sim -1$, which deviates from the prediction of the SM by $4.3\,\sigma$~\cite{Altmannshofer:2017fio, Altmannshofer:2015sma}. In contrast, the Wilsonian coefficient $C^\mu_{10}$ agrees with the prediction of the SM, which can be determined to a remarkable precision by the well measured quantity BR$(B_s \to \mu^+\mu^-)$.

New physics are introduced to explain these anomalies observed in $b\to s$ rare transitions. For example, by introducing new operators in the effective Hamiltonian, model independent fits \cite{Descotes-Genon:2013vna,Hurth:2014vma,Altmannshofer:2014rta,Hiller:2014yaa} 
have been considered. It is found that the NP operators in the form $(\bar{s}\gamma_\mu P_L b)(\bar{\ell} \gamma^\mu P_L \ell)$ can be consistent with the explanations for the $B\to K^{(*)} \mu^- \mu^+$ angular distributions measured by the LHCb collaboration. $Z^\prime$ models are considered in Refs.~
\cite{Descotes-Genon:2013wba,Gauld:2013qba,Buras:2013qja,Gauld:2013qja,Buras:2013dea, Altmannshofer:2014cfa,Crivellin:2015mga,Crivellin:2015lwa,Niehoff:2015bfa,Sierra:2015fma,Celis:2015ara}
and leptoquark models are examined in
Refs.~\cite{Becirevic:2015asa,Varzielas:2015iva}. Furthermore, it has been
argued that as the violation of lepton flavour universality violation in $R_K$
as well as in $B$ decays~\cite{Glashow:2014iga} might be linked to neutrino
oscillations \cite{Boucenna:2015raa}. \\
In this work, we will explore these anomalies in the context of type III Two-Higgs Doublet model (2HDM). There are several studies on $B\to K^{(*)} \mu^- \mu^+$  in the type II 2HDM \cite{Bobeth:2001sq,Jung:2012vu,Hu:2016gpe, Arnan:2017lxi} and it was found that the type-II 2HDM could not explain the anomaly of $R_{K^{(*)}}$ in the current world average.


As a minimal extension of the SM scalar sector, the scalar spectrum of 2HDM
(consists of two charged Higgs $H^\pm$, one CP-odd $A$, and two CP-even $h$
and $H$ (one of them can be identified as SM-like Higgs boson found at the
LHC). 
This model can accommodate the electroweak test precision data, B physics
data, and Higgs data as well. Complementary to direct searches, indirect
constraints on the general 2HDM could be obtained from the rare FCNC decays,
since Higgs bosons in this model can affect these processes through the
penguin and box diagrams. Typically, the most general version of 2HDM has
non-diagonal fermionic couplings in flavor space, and can therefore generate
tree-level flavor-changing neutral current (FCNC) phenomena, which might be
inconsistent with observed data. Several ways to suppress FCNCs have been
suggested in the literature. The simplest one is to impose $Z_2$ symmetry
which forbid unwanted non-diagonal terms. Depending on the $Z_2$ charge
assignments to the scalars and fermions, it results in four types of 2HDMs
(types, I, II, X, Y)\cite{Branco:2011iw}. An alternative solution is to assume
the so-called Cheng-Sher ansatz in the fermion sector which force the
non-diagonal Yukawa couplings to proportional to the mass of the involved
fermions, i.e $Y_{ij} \varpropto \sqrt{m_i m_j}/v$, which is called Type-III
2HDM \cite{Cheng:1987rs}. In this scenario, the absence of tree-level FCNCs is
automatically guaranteed by assuming the alignment in flavor space of the
Yukawa matrices. In this work, the type-III 2HDM will be carefully
examined. We find the parameter region with large $\tan\beta$ (say $30 <
\tan\beta < 50$) and light charged Higgs boson (say $150$ GeV $< m_{h^\pm} < 350$ GeV) can offer an explanation to the measured $R_{K}$ value and accommodate pretty well the other B physics data, like  $\Delta M_q$ ($q=s,d$), $B \to  X_s \gamma$, $B_q \to \mu^+ \mu^{-}$, and $B_q \to X_s \mu^+ \mu^{-}$.

The paper is organized as follows: In section~\ref{sec:formalism}, we review the Yukawa sector in the type-III 2HDM. In section~\ref{sec:flavourobservable} we study constraints from $B^-_{q''} \to \tau \bar\nu$, $ B_q-\bar B_q$ mixing, and $\bar B\to X_s \gamma$ followed by constraints from $B_{q}\to \mu^+ \mu^-$ and $B_q \to X_s \mu^+ \mu^-$ in section \ref{sec:BsmmandBsmm}. In section~\ref{sec:RKRKs}, we examine the results of $R_K$ and $R_{K^{(*)}}$ in type III of 2HDM. In section \ref{sec:conlusion} we summarize our studies.

\section{ Yukawa sector in the Type-III 2HDM }
\label{sec:formalism}
In this section, we briefly describe the Yukawa sector of the type-III 2HDM. 
In order to derive the scalar Yukawa couplings to the SM quarks and leptons, we put the Higgs doublets $\Phi_{1}$ and $\Phi_{2}$ as:
\begin{align}
\Phi_i=
\begin{pmatrix}
\omega_i^+\\
\frac1{\sqrt2}(v_i + h_i + i z_i)
\end{pmatrix} \,, \label{eq:Phi_i}
\end{align}
 where there are eight scalar fields, $v_{1(2)}$ is the vacuum expectation value (VEV) of $\Phi_{1(2)}$, which is related to the $W$-boson mass as $m_W=g v/2$ with $v=\sqrt{v^2_1 + v^2_2} \approx 246$ GeV ($g$ being the $SU(2)_L$ gauge coupling). After the spontaneous symmetry breaking $SU(2)_L \times U(1)_Y \to U(1)_{em}$, three of eight scalar fields become pseudo-Nambu-Goldstone bosons. The remaining five scalar fields are physical states, which include two charged-Higgs $(H^\pm)$, 
one CP-odd pseudoscalar $(A)$, and two CP-even scalars $(H, h)$. Accordingly, the physical and weak eigenstates can be expressed as:
  \begin{equation}
\begin{pmatrix}h_1\\h_2\end{pmatrix}=R(\alpha)
\begin{pmatrix}H\\h\end{pmatrix},\quad
\begin{pmatrix}z_1\\z_2\end{pmatrix}=R(\beta)
\begin{pmatrix}z\\A\end{pmatrix},\quad
\begin{pmatrix}\omega_1^+\\\omega_2^+\end{pmatrix}=R(\beta)
\begin{pmatrix}\omega^+\\H^+\end{pmatrix}\,,   \label{eq:rots}
 \end{equation}
where angle $\alpha$ denotes the mixing between the two CP-even $H$ and $h$; angle $\beta$ is defined by $\cos\beta (\sin\beta)=v_{1(2)}/v$; $z$ and $\omega^\pm$ denote the Nambu-Goldstone bosons, and the three rotating matrices can be unified as:
 \begin{align}
R(\theta)=\begin{pmatrix}\cos\theta&-\sin\theta\\
\sin\theta&\cos\theta\end{pmatrix}.
\end{align}
For the purpose of phenomenological study, it is convenient to set $\sin(\beta-\alpha) = 1$. 
\subsection{Neutral scalar Yukawa couplings}
In the type-III 2HDM, the Yukawa couplings to the quarks and leptons can be written as:
\begin{align}
- \mathcal L_Y=
\bar Q_L( Y_1^d\Phi_1+Y_2^d\Phi_2) d_R
+\bar Q_L(Y_1^u \tilde\Phi_1+Y_2^u \tilde\Phi_2)u_R
+\bar L_L(Y_1^\ell\Phi_1+Y_2^\ell\Phi_2) e_R
+\text{H.c.},
\end{align}
where the flavor indices are suppressed; $\tilde\Phi_i=i\sigma_2\Phi_i^*$ and  $\sigma_2$  is the Pauli
matrix; $Q_L (L_L)$ denotes the left-handed doublet quarks (leptons); $u_R$, $d_R$, and $e_R$ are the right-handed up-type quarks, down-type quarks, and charged-leptons, respectively,   and $Y_i^f$ are the $3\times3$ complex Yukawa matrices in flavor space. Using Eq.~(\ref{eq:Phi_i}), the fermion mass matrix can be formulated as:
 \begin{equation}
-{\cal L}_{Y} \supset  \bar f_L {\bf M}^f f_R + H.c. \equiv  \bar f_L \left( \frac{ v_1 Y^f_1}{\sqrt{2} } + \frac{ v_2 Y^f_2}{\sqrt{2} } \right) f_R + H.c. 
 \end{equation}
Without assuming the relation between $Y^f_1$ and $Y^f_2$, in general, both Yukawa matrices cannot be simultaneously diagonalized which leads to the flavor-changing neutral currents (FCNCs) mediated by scalar bosons at the tree level. 

To diagonal fermion mass matrix, we introduce the unitary matrices $U^f_L$ and $U^f_R$, where the physical and weak states are related by $f^p_L = U^f_L f^w_L$ and  $f^p_R = U^f_R f^w_R$. Thus, the  couplings of neutral scalars to quarks can be expressed as:
 \begin{align}
 -{\cal L}^\phi_Y & = \bar u_L \left[ \left( \frac{c_\alpha {\bf m}^u}{v s_\beta} - \frac{c_{\beta-\alpha} \Xi^u}{\sqrt{2} s_\beta}\right) h + \left( \frac{s_\alpha {\bf m}^u}{ s_\beta v} + \frac{s_{\beta-\alpha} \Xi^u}{\sqrt{2} s_\beta}\right) H \right] u_R \nonumber \\
 &+ \bar d_L \left[ \left( - \frac{s_\alpha {\bf m}^d}{v c_\beta} + \frac{c_{\beta-\alpha} \Xi^d}{\sqrt{2} c_\beta}\right)h + \left( \frac{c_\alpha {\bf m}^d}{ c_\beta v} - \frac{s_{\beta-\alpha} \Xi^d}{\sqrt{2} c_\beta}\right) H \right] d_R \nonumber \\
 %
 %
& - i \left[ \bar u_L \left( \frac{ {\bf m}^u}{ t_\beta v} - \frac{ \Xi^u }{\sqrt{2} s_\beta}\right) u_R  
 + \bar d_L \left( \frac{ t_\beta{\bf m}^d}{ v} - \frac{ \Xi^d }{\sqrt{2} c_\beta}\right) d_R 
 \right] A + H.c. \,, \label{eq:Yu_phi}
 \end{align}
where ${\bf m}^f = U^f_L {\bf M}^f U^{f\dagger}_R$ denotes the diagonal mass matrix, $\Xi^{u} = U^u_L Y^u_1 U^{u\dagger}_R$, $\Xi^{d} = U^d_L Y^d_2 U^{d\dagger}_R$, $c_\alpha(s_\alpha) =\cos\alpha (\sin\alpha)$, $c_\beta(s_\beta)=\cos\beta(\sin\beta)$, $c_{\beta-\alpha}[s_{\beta-\alpha}]=\cos(\beta-\alpha) [\sin(\beta-\alpha)]$, and $t_\beta=\tan\beta$. We note that the couplings of the charged-leptons can be 
obtained in a straightforward way when ${\bf m}^d$, $Y^d_2$, and $U^d_{L,R}$ are replaced by ${\bf m}^\ell$, $Y^\ell_2$, and $U^\ell_{L,R}$, respectively.  
From Eq.~(\ref{eq:Yu_phi}),  it is noticed that tree-level FCNC processes  are associated with non-vanishing $\Xi^u$, $\Xi^d$, and $\Xi^\ell$; and when they vanish, it can be realized either by imposing the 
alignment of the two Yukawa matrices, the Yukawa interactions are returned to the type-II 2HDM.  

In order to naturally suppress the FCNCs at the tree level, we adopt the so-called Cheng-Sher ansatz~\cite{Cheng:1987rs} in the quark and lepton sectors, where $\Xi^f$ is  parameterized as $\Xi^f_{ij} = \sqrt{m^f_i m^f_j} \chi^f_{ij} /v$, and $\chi^f_{ij}$ are taken as dimensionless free parameters.  Although in general $\chi^f_{ij} \neq \chi^f_{ji}$ with $i \neq j$,  to simplify the numerical analysis, we assume $\chi^f_{ij} = \chi^f_{ji}$ in our analysis.  Thus, the  neutral scalar Yukawa couplings to the quarks and leptons can be generally written as:
\begin{equation}
-{\cal L}^\phi_Y= \sum_{f=u,d,\ell} \frac{m^f_j }{v} \left[ (\xi^f_h)_{ij}  \bar f_{Li}  f_{Rj}  h + (\xi^f_H)_{ij} \bar f_{Li}  f_{Rj} H - i (\xi^f_A)_{ij} \bar f_{Li}  f_{Rj} A \right] + H.c.\,,
\label{Eq:Yukawa}
\end{equation}
where $(\xi^f_{\phi})_{ij}$ with $\phi=h, H, A$ are given in Table~\ref{tab:MixFactor}. 

\begin{table}[!h]
\caption {Yukawa couplings of the $h$, $H$, and $A$ bosons to the quarks and leptons in type-III 2HDM.  The couplings in type-II 2HDM can be easily obtained when $\chi^f_{ij}$ vanish. } 
\label{tab:MixFactor}
\begin{center}
\begin{tabular}{c|c|c|c} \hline\hline 
$\phi$  & $(\xi^u_{\phi})_{ij}$ &  $(\xi^d_{\phi})_{ij}$ &  $(\xi^\ell_{\phi})_{ij}$  \\   \hline
$h$~ 
        & ~ $  \frac{c_\alpha}{s_\beta} \delta_{ij} -  \frac{c_{\beta-\alpha}}{\sqrt{2}s_\beta}  \sqrt{\frac{m^u_i}{m^u_j}} \chi^u_{ij}$~
        & ~ $ -\frac{s_\alpha}{c_\beta} \delta_{ij} +  \frac{c_{\beta-\alpha}}{\sqrt{2}c_\beta} \sqrt{\frac{m^d_i}{m^d_j}}\chi^d_{ij}$~
        & ~ $ -\frac{s_\alpha}{c_\beta} \delta_{ij} + \frac{c_{\beta-\alpha}}{\sqrt{2}c_\beta} \sqrt{\frac{m^\ell_i}{m^\ell_j}}  \chi^\ell_{ij}$ ~ \\
$H$~
        & $ \frac{s_\alpha}{s_\beta} \delta_{ij} + \frac{s_{\beta-\alpha}}{\sqrt{2}s_\beta} \sqrt{\frac{m^u_i}{m^u_j}} \chi^u_{ij} $
        & $ \frac{c_\alpha}{c_\beta} \delta_{ij} - \frac{s_{\beta-\alpha}}{\sqrt{2}c_\beta} \sqrt{\frac{m^d_i}{m^d_j}}\chi^d_{ij} $ 
        & $ \frac{c_\alpha}{c_\beta} \delta_{ij} -  \frac{s_{\beta-\alpha}}{\sqrt{2}c_\beta} \sqrt{\frac{m^\ell_i}{m^\ell_j}}  \chi^\ell_{ij}$ \\
$A$~  
        & $ \frac{1}{t_\beta} \delta_{ij}- \frac{1}{\sqrt{2}s_\beta} \sqrt{\frac{m^u_i}{m^u_j}} \chi^u_{ij} $  
        & $ t_\beta \delta_{ij} - \frac{1}{\sqrt{2}c_\beta} \sqrt{\frac{m^d_i}{m^d_j}}\chi^d_{ij}$  
        & $t_\beta \delta_{ij} -  \frac{1}{\sqrt{2}c_\beta} \sqrt{\frac{m^\ell_i}{m^\ell_j}}  \chi^\ell_{ij}$ \\ \hline \hline 
\end{tabular}
\end{center}
\end{table}
\subsection{Yukawa couplings of Charged Higgs Boson}
The rotation matrix for charged scalars in Eq.~(\ref{eq:rots}) is the same as 
that  for pseudoscalars; therefore, the  Yukawa couplings of the charged Higgs boson are similar to those of the CP-odd scalar and can be written as: 
\begin{align}
{\cal L}^{H^\pm}_Y & = \frac{\sqrt{2}}{v} \bar u_{i }  \left( m^u_i  (\xi^{u*}_A)_{ki}  V_{kj} P_L + V_{ik}  (\xi^d_A)_{kj}  m^d_j P_R \right) d_{j}  H^+  \nonumber \\
& + \frac{\sqrt{2}}{v}  \bar \nu_i  (\xi^\ell_A)_{ij} m^\ell_j P_R \ell_j H^+ + H.c.\,, \label{eq:Yukawa_CH}
\end{align}
where the sum over flavor indices is indicated, $V\equiv V^u_L V^{d\dagger}_L$ is the Cabibbo-Kobayashi-Maskawa (CKM) matrix, and $P_{R,L}=(1 \pm \gamma_5)/2$ are the chiral projection operators.  Since the CKM matrix elements have hierarchy properties when different generations of fermions are involved, in the following, we examine the possible enhancement factor for $\bar u_i b H^+$ and $\bar u_i s H^+$ vertices in the type-III model. For the sake of convenience, we define $C^L_{ij} = m^u_i  (\xi^{u*}_A)_{ki}  V_{kj}$ and $C^R_{ij} = V_{ik}  (\xi^d_A)_{kj}  m^d_j$. \\ 
  \underline {$  u b H^+$ vertex:} With $\xi^{u,d}_{A}$ shown in Table~\ref{tab:MixFactor} and $t_\beta > 1$, the $C^L_{ub}$ coupling can be simplified as: 
 \begin{equation}
 C^L_{ub} = m_u \left( \frac{1}{t_\beta} - \frac{\chi^u_{11} }{\sqrt{2} s_\beta}  \right)V_{ub} - \frac{\sqrt{m_u m_c} \chi^{u}_{21}}{\sqrt{2} s_\beta} V_{cb} - \frac{\sqrt{m_u m_t} \chi^{u}_{31}}{\sqrt{2} s_\beta} V_{tb} \approx - \frac{\sqrt{m_u m_t} \chi^{u}_{31}}{\sqrt{2} s_\beta} V_{tb}\,. \label{eq:CLub}
 \end{equation}
It can be seen that due to $O(\sqrt{m_u m_t}/v) \sim V_{ub}$, unless $\chi^{u}_{31}\ll 1$, $C^L_{ub}$  can  has a sizable effect on the $b\to u$ decay, where the one in type-II is negligible. 
The situations in $C^R_{ub}$ are different from $C^L_{ub}$. If we decompose $C^R_{ub}$ to be:
  \begin{equation}
 C^R_{ub} =- V_{ud}  \frac{\sqrt{m_d m_b} \chi^d_{13} }{\sqrt{2} c_\beta} -  V_{us} \frac{\sqrt{m_s m_b} \chi^{d}_{23}}{\sqrt{2} c_\beta} + V_{ub} m_b \left(  t_\beta- \frac{ \chi^{d}_{33}}{\sqrt{2} c_\beta} \right)  \,, \label{eq:CRub}
 \end{equation}
it can be seen that due to $V_{ud} \sqrt{m_d m_b} \sim V_{us} \sqrt{m_s m_b} \gg  V_{ub} m_b$, the first two  terms in Eq.~(\ref{eq:CRub}) are compatible and cannot be neglected. However, if we further assume $\chi^d_{13, 23} \ll 0.1$, we then have $C^R_{ub} \approx V_{ub} (\xi^d_A)_{33} m_b$, which is similar to the coupling in the type-II case.  \\
\underline {$ c(t) b H^+$ vertex:}  Following the above discussions for the $ubH^+$ coupling,  the $C^L_{cb}$ coupling can be expressed as:
   \begin{equation}
    C^L_{cb} = m_c (\xi^{u*}_A)_{12} V_{ub} + m_c (\xi^{u*}_A)_{22} V_{cb} + m_c (\xi^{u*}_A)_{32} V_{tb}\approx - \frac{\sqrt{ m_c m_t} }{\sqrt{2} s_\beta}\chi^{u}_{32} V_{tb}\,, \label{eq:CLcb}
    \end{equation} 
  where  $V_{tb} \gg V_{cb}\gg V_{ub}$ is used.
     It is of interest to  numerically see $\sqrt{m_c m_t}/v \sim V_{cb}$; that is, if $\chi^u_{32}$ or $\chi^u_{tc}$ is of $O(1)$, the charged-Higgs effect $C^L_{cb}$ will significantly enhance the $b\to c$ decays.   
Based on the fact that $|V_{cd}| \sqrt{m_d m_b}\ll$  $V_{cs}$ $\sqrt{m_s m_b}$ and $V_{cb} m_b$, $C^R_{cb}$ can be written as:
   \begin{equation}
   C^R_{cb} 
    \approx - V_{cs} \frac{\sqrt{m_s m_b}\chi^d_{23} }{\sqrt{2} c_\beta} + V_{cb}m_b \left( t_\beta - \frac{\chi^d_{33}}{\sqrt{2} c_\beta}\right)\,, \label{eq:CRcb}
   \end{equation}
   where due to $V_{cb} m_b < V_{cs} \sqrt{m_s m_b}$, the first term in
   $C^R_{cb}$ cannot be neglected except if $\chi^d_{23} \ll 0.1$.  Since
   $t$-$b$-$H^+$ couplings are normally associated with large CKM matrix element with $V_{tb} \approx 1$, therefore, they can be expressed as: $C^L_{tb} \approx m_t (\xi^{u*}_A)_{33} V_{tb}$ and $C^R_{tb} \approx V_{tb} (\xi^{d}_A)_{33} m_b$. 
 \noindent
\underline {$ u(c) s H^+$ vertex:}  To analyze the $u(c)$-$s$-$H^+$ couplings, it is convenient to include the factor $\sqrt{2}/v$. Thus, $C^{L,R}_{us}$ and $C^{L,R}_{cs}$ can be reduced to be:
  \begin{align}
  \frac{\sqrt{2}}{v} C^L_{us} & \approx - \frac{\sqrt{m_u m_c}}{s_\beta v} \chi^u_{21} V_{cs} -  \frac{\sqrt{m_u m_t}}{s_\beta v} \chi^u_{31} V_{ts}  \ll 1 \ \  \text{ \rm (negligibly)} \nonumber \\
  %
   %
   \frac{\sqrt{2}}{v} C^L_{cs} & \approx  \frac{\sqrt{2} m_c }{v } (\xi^{u*}_A)_{22} V_{cs} -  \frac{\sqrt{m_c m_t}}{s_\beta v} \chi^u_{32} V_{ts}  \ll 1 \ \  \text{ \rm (negligibly)} \nonumber \\
  \frac{\sqrt{2}}{v} \frac{C^R_{us}}{V_{us}} & \approx  \frac{\sqrt{2}}{v}\frac{ C^R_{cs}}{V_{cs}}  \approx \frac{\sqrt{2} m_s}{v}  \left( t_\beta - \frac{\chi^d_{22} }{\sqrt{2} c_\beta} \right)\,.
  \end{align}
Although $m_s/v \sim 3.9 \times 10^{-4}$  in $C^R_{us, cs}/v$ is a suppression factor, due to an enhancement from a large $t_\beta$ or $1/c_\beta$,  $u(c)$-$s$-$H^\pm$ coupling can reach a few percent  level. Since there is no other enhancement factor in $C^L_{us(cs)}$, their couplings are below $1\%$ and can be neglected. 
 \noindent
 \underline {$ t s H^+$ vertex:}   $m_t V_{ts} \sim 6.72 \ {\rm GeV} < \sqrt{m_c m_t} V_{cs} \sim 14.8$ GeV,  $ m_s V_{ts} \ll \sqrt{m_s m_b} V_{tb} \sim 0.66$ GeV, we can simplify $C^{L,R}_{ts}$ to be:
  \begin{align}
  C^L_{ts} & \approx - \frac{\sqrt{m_c m_t }}{\sqrt{2} s_\beta} \chi^u_{23} V_{cs} + m_t  \left( \frac{1}{t_\beta} - \frac{\chi^u_{33} }{\sqrt{2} s_\beta}\right) V_{ts} \,, \label{eq:CLts} \\
  C^R_{ts} & \approx  - V_{tb} \frac{\sqrt{m_s m_b}}{\sqrt{2} s_\beta} \chi^d_{32} \,. \nonumber 
  \end{align}
 If $\chi^u_{23} > \chi^u_{33}$, numerically, we can drop the second term in $C^L_{ts}$, which  only involves $\chi^u_{23}$. 

\section{ Constraints from $B^-_{q''} \to \tau \bar\nu$, $ B_q-\bar B_q$ mixing, and $\bar B\to X_s \gamma$}
\label{sec:flavourobservable}

From the discussions in section.2, the essential ingredients in the Yukawa sector  
especially the couplings of the charged Higgs scalars to quarks and leptons are extracted. Obviously, the new free parameters are associated to the masses of quarks in the leading contributions. 
Consequently, we argued that the lightest charged Higgs with the new couplings might have interesting phenomenologies in some rare decays which are suppressed in the SM. 
Hence, in the following analysis, we will focus on the contributions of charged Higgs boson as well as neural Higgs bosons to the relevant FCNC processes of B mesons. 

  
\subsection{Limits from $B^-_{q''} \to \tau \bar \nu_\tau$ ($q''=u,c$)}
\begin{figure}[phtb]
\hspace{-2.55cm}
\includegraphics[scale=0.9]{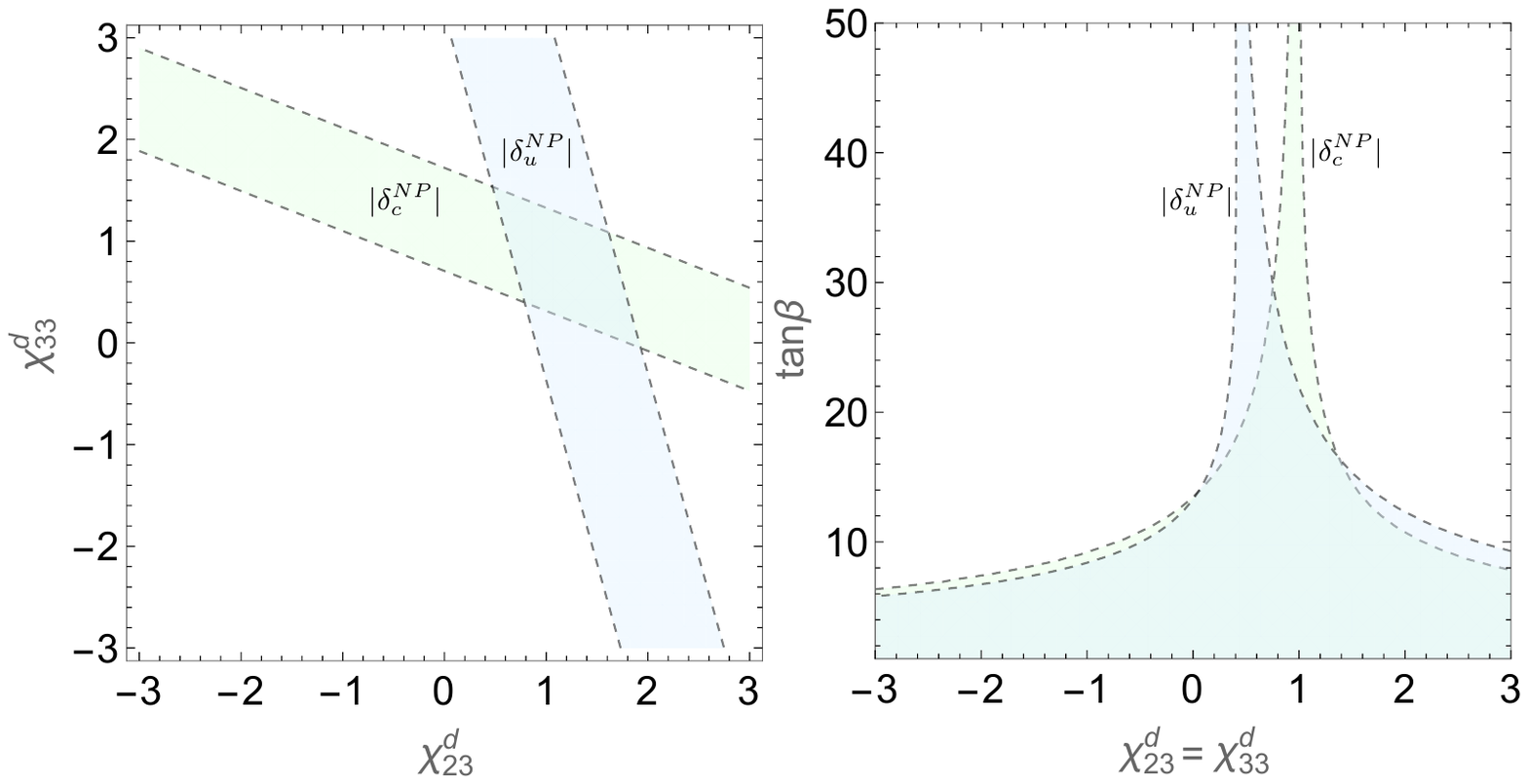}
\vspace{-18.1cm}
\caption{Allowed ranges of $|\delta^{NP}_u|$ (light green) and $|\delta^{NP}_c|$ (light blue) in the ($\chi^{d}_{23}, \chi^{d}_{33}$)(left) and ($\chi^{d}_{23}, \tan\beta $)(right) planes,  We use $m_{H^\pm}=200$ GeV in both plots. }
\label{fig:dNP}
\end{figure}

As emphasized in section.2, we know that a CKM suppression charged-Higgs
coupling in type-II model can be turned to a CKM enhancement coupling in
type-III model. Since the CKM matrix elements  are well measured in
experiments, a rare decay process may give a stringent constraint on the
$\chi^{u,d}_{ij}$ new parameters when a large CKM is involved in the 
interaction vertex.  To understand the constraints, we consider the 
$B_{u(c)} \to \tau \bar \nu_\tau$ decays, where the branching ratio (BR) 
for $B_u \to \tau \bar \nu_\tau$ averaged by heavy flavor averaging group
(HFAG) is $BR(B_u \to \tau \bar\nu_\tau) = (1.06 \pm 0.19)\times
10^{-4}$. Although $B_c \to \tau \bar \nu_\tau$ has not yet been observed,
using the difference in $B_c$ lifetime between the SM and experimental
results, the upper limit is obtained as  $BR(B_c \to \tau \bar \nu_\tau) <
30\%$~\cite{Alonso:2016oyd}. 
From Eqs.~(\ref{eq:CLub}),(\ref{eq:CRub}),(\ref{eq:CLcb}), and (\ref{eq:CRcb}), it can be seen that each interaction vertex may  involve several parameters; in order to understand the effects of each parameter, when we focus on one term in $C^{L(R)}_{q'' b}$, we will turn off the contributions from the others if the vertex consists of one more different Yukawa coupling. In addition, the charged-Higgs couplings to the leptons also involve new free parameters $\chi^\ell_{ij}$, which are completely independent of $\chi^{u(d)}_{ij}$; 
Thus, the charged-Higgs couplings used in this section are expressed as:
   \begin{align}
   {\cal L}^{H^\pm}_Y&  \supset   \frac{\sqrt{2} }{v}\bar q'' \left[ \left( C^L_{q'' b} P_L + C^R_{q'' b} P_R \right) b + m_\tau t_\beta  \bar \nu_\tau P_R \tau\right] H^+ + H.c.\,, 
   \end{align}
 where $C^{L,R}_{q'' b}$ can be found from Eqs.~(\ref{eq:CLub}-\ref{eq:CRcb}). Accordingly, the BR for $B_{q''} \to \tau \bar\nu_\tau$ can be given as:
  \begin{align}
  BR(B_{q''} \to \tau \bar \nu_\tau) & = BR^{\rm SM}(B_{q''} \to \tau \bar \nu_\tau) \left| 1 - \frac{ (C^R_{q'' b} -C^L_{q'' b})  m^2_{B_{q}} t_\beta}
{V_{q'' b}  (m_{q'' } + m_b) m^2_{H^\pm} }\right|^2\,\,, \\
  BR^{\rm SM}(B_{q'' } \to \tau \bar \nu_\tau) &= \frac{G^2_F |V_{q'' b}|^2}{8 \pi}  f^2_{B_{q'' }} m_{B_{q'' }} m^2_\tau \left(1- \frac{m^2_\tau}{m^2_{B_{q'' }}} \right)^2 \,. \nonumber 
  \end{align}
Using $V_{ub} \approx 3. 72 \times 10^{-3} e^{-i\gamma}$ with $\gamma\approx 70^{\circ}$, $V_{cb} \approx 0.04$~\cite{PDG}, $m_{B_{u(c)}}\approx 5.28 (6.27)$ GeV, $f_{B_u} =0.191$ GeV~\cite{Lenz:2010gu}, and $f_{B_c}=0.434$ GeV~\cite{Colquhoun:2015oha}, we obtain $BR^{\rm SM}(B_u \to \tau \bar \nu_\tau) = 0.89\times 10^{-4}$ and $BR^{\rm SM}(B_c \to \tau \bar \nu_\tau)\approx 0.02$. Since $BR^{\rm exp}(B_u \to \tau \bar \nu_\tau)/BR^{\rm SM}(B_u \to \tau \bar \nu_\tau) \sim 1.19$, if the new physics effect is required to be smaller than the SM contribution and to be within $1\sigma$ errors of data,  the free parameter can be limited as:
\begin{equation}
  |\delta^{NP}_{q'' }| = \left|  \frac{(C^R_{q'' b} - C^L_{q'' b}) m^2_{B_{q'' }} t_\beta }{V_{q'' b} (m_{q'' } + m_b) m^2_{H^\pm}} \right|  \leq  \left\{\begin{array} {c} 
  0.1  \  \text{ ($q'' =u$)}\,,  \\
  4.0 \ \    \text{($q'' =c$)} \,. \end{array} \right. \,,  
 \end{equation}
\textcolor{black} {where the $ |\delta^{NP}_c|$ upper bound is from the result of $BR(B_c \to \tau \bar \nu_\tau)< 30\%$.  Accordingly, we show  $|\delta^{NP}_{u}|$ in the ($\chi^{d}_{23}, \chi^{d}_{33}$) in left plot of Fig.~\ref{fig:dNP}, where we assume $\chi^u_{31}$ is less than $1\%$ through the $B_u \to \tau \bar \nu_\tau$ measurement. Since the upper bound from the $B_c \to \tau \bar \nu_\tau$ decay is still much larger than the SM result, $\chi^u_{32}$ can still be of $O(1)$. In the right panel of Fig.~\ref{fig:dNP} we show the allowed regions in the ($\chi^d_{23}, \tan\beta$) plane, as it can be seen large $\tan\beta$ is preferred when $\chi^{d}_{33, 23} \sim O(1)$. In both plots, we use  $m_{H^\pm}=200$ GeV.
}
\subsection{ Constraints from $\Delta M_q$ }
\begin{figure}[phtb]
\includegraphics[scale=0.85]{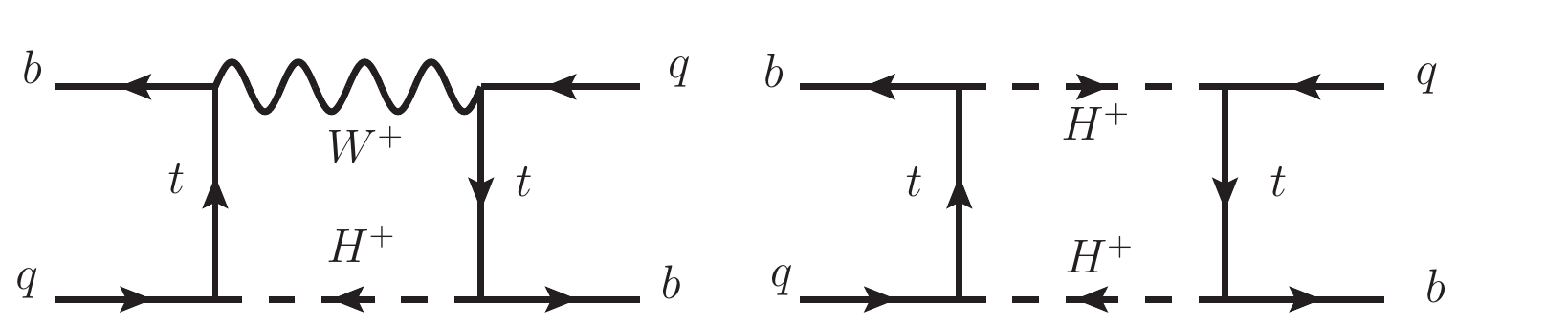}
\caption{The sketched box diagrams for the $ B_{q}- \bar{B}_{q}$ mixing mediated by the $W^+$ and $H^+$ bosons.}
\label{fig:boxes}
\end{figure}

Let us now consider the bounds from the $\Delta B=2$ processes. 
It is known that the tree-level FCNCs can be induced by the generic 2HDM,
therefore, the measured $\Delta M_{q}$ $(q=d,s)$ usually gives a strict limit
on the  parameters $\Xi^q_{ij}$. However, due to the suppression of
$\sqrt{m^q_i m^q_j}/v$ from the Cheng-Sher ansatz, $\Delta B=2$ processes
mediated by the neutral scalars at the tree level are small and negligible.
Therefore, the main contributions to the $\Delta B=2$ processes in this study
are still from the box diagrams, which arise from the $W^\pm$ and $H^\pm$
bosons, where  the typical Feynman diagrams in 2HDM mediated by $W^\pm$ -
$H^\mp$ and $H^\pm$-$H^\mp$ are sketched in Fig.~\ref{fig:boxes}.  In
addition,  the  Yukawa couplings of $H^\pm$ to the quarks are proportional to
the quark masses. Thus, the heavier the quarks are, the lager is the 
enhancement of the  $H^\pm$ effects. Hence, we only consider the top-quark loop contributions in B-meson system. The relevant charged-Higgs interactions are given as:
 \begin{equation}
 {\cal L}^{H^\pm}_Y \supset \frac{\sqrt{2}}{v} V_{tb} \bar t \left( m_t \zeta^u_{tt} P_L + m_b \zeta^d_{bb} P_R \right) b  H^+ + \frac{\sqrt{2}}{v} V_{tq} \bar t  \left( m_t \zeta^u_{tq} P_L  \right) q  H^+ +H.c.,  \label{eq:btoqpp}
 \end{equation}
where  using the scheme $\chi^d_{ij} \approx 0$ ($i\neq j$), Eq.~(\ref{eq:CLts}), and $m_{q} \approx 0$, the coefficients $\zeta^{q'}_{ij}$ are given as:
\begin{align}
\zeta^u_{tt} & \approx  \frac{1}{t_\beta} \left(1- \frac{\chi^u_{33} }{\sqrt{2} c_\beta} \right)=\frac{z^u_{33}}{t_\beta}  \,, \  \ \zeta^d_{bb}  \approx  t_\beta \left( 1- \frac{\chi^d_{33}}{\sqrt{2} s_\beta}\right) = t_\beta z^d_{33}\,, \nonumber \\
\zeta^u_{tq} & \approx  \frac{1}{t_\beta} \left(1- \frac{\chi^u_{33} }{\sqrt{2} c_\beta} - \frac{\chi^u_{23} }{\sqrt{2} c_\beta} \sqrt{\frac{m_c}{m_t}} \frac{V_{cq}}{V_{tq}}\right) = \frac{z^u_{23}}{t_\beta}  \,.  \label{eq:zetas}
\end{align}
From Eqs.~(\ref{eq:btoqpp}) and (\ref{eq:zetas}), it can be clearly seen that when $z^u_{33}=z^d_{33}=z^u_{23}=1$,  the type-II 2HDM is reproduced. Unlike the type-II model, $z^u_{33, 23}\gg 1$ can be achieved in type-III model to compensate the suppression of $1/t_\beta$ at large values of $t_\beta$. It is of interest to mention that due to the enhancement $|V_{cq}/V_{tq}|$, the third term, which has the factor $\sqrt{m_c/m_t} V_{cq}/V_{tq} \sim 2$, in $\zeta^u_{tq}$ is not suppressed and can be comparable with the second term. These new  2HDM effects may have important impacts on the flavor physics, which we want to explore in this  work. 
Based on the conventions in~\cite{Becirevic:2001jj}, the effective Hamiltonian is written as:
 \begin{equation}
 H^{\Delta B=2}_{\rm eff} = \frac{ G^2_F \left( V^*_{tb} V_{tq}\right)^2 }{16 \pi^2} m^2_W  \sum_i  C_i (\mu) Q_{i} \,, 
 \end{equation}
where the effective operators are given as:
 %
 %
 \begin{eqnarray}
        Q_1 & = & (\bar{d}^{\alpha}_L \gamma_\mu b^{\alpha}_L)( \bar{d}^{\beta}_{L} \gamma^\mu
        b^{\beta}_L)
        \nonumber \\
        Q_2 & = & (\bar{d}^{\alpha}_R  b^{\alpha}_L)( \bar{d}^{\beta}_R b^{\beta}_L)\ 
        \nonumber \\
        Q_3 & = & (\bar{d}^{\alpha}_R  b^{\beta}_L)( \bar{d}^{\beta}_R b^{\alpha}_L)\ 
       \\
        Q_4 & = & (\bar{d}^{\alpha}_R  b^{\alpha}_L) (\bar{d}^{\beta}_L b^{\beta}_R)\ 
        \nonumber \\
        Q_5 & = & (\bar{d}^{\alpha}_R  b^{\beta}_L ) (\bar{d}^{\beta}_L b^{\alpha}_R)\ 
       \nonumber
\end{eqnarray}
with $\alpha, \beta$ being the color indices. The Wilson coefficients at the scale $\mu=m_b=4.8$ GeV can be expressed as~
\cite{Becirevic:2001jj}:
 \begin{equation}
 C_i(m_b) \approx \sum_{k,j} \left( b^{(i,j)}_k + \eta c^{(i,j)}_k \right) \eta^{a_k} C_j(\mu_H)\,, \label{eq:mbWCs}
 \end{equation}
 where $\mu_H = m_{H^\pm}$, $\eta = \alpha_s(\mu_H)/\alpha_s(m_t)$, $C_j(\mu_H)$ are the Wilson coefficients at $\mu_H$ scale, and the magic numbers of $a^{i,j}_k$, $b^{i,j}_k$, and $c^{i,j}_k$ can be found in~\cite{Becirevic:2001jj}.   The non-vanishing Wilson coefficients at the $\mu_H$ scale induced from the $WW$, $WH$, and $HH$ diagrams shown in Fig.~\ref{fig:boxes} are $C_1(\mu_H) = C^{\rm SM}_1 + C^{WH}_1 + C^{HH}_1$ and $C_2(\mu_H)=C^{\rm HH}_2$, where the SM result is $C^{\rm SM}_1 = 4 S_0(m^2_t/m^2_W)=3.136 (m^2_t/m^2_W)^{0.76} \approx 9.36$~\cite{Buchalla:1995vs}, and $S_0(x)$ is the Inami-Lin function~\cite{Inami:1980fz}.  If we define  $x_t=m^2_t/m^2_W$, $y_t=m^2_t/m^2_{H^\pm}$, $y_W=m^2_W/m^2_{H^\pm}$, and $y_b = m^2_b/m_H^\pm$,  the results from the charged-Higgs contributions can be formulated as:
\begin{align}
 %
 C^{WH}_{1} &= 2 \zeta^u_{tq} \zeta^{u*}_{tt} y_t \left\{ \frac{y_t (4-x_t)}{(1-y_t)(y_t -y_W)} + \frac{y_W (4 y_t - y_W x_t)}{(1-y_W) (y_t - y_W)^2} \ln(y_W)\right. \nonumber \\
  & \left.  - \frac{y_t y_W}{(1-y_t)^2 (y_t -y_W)^2 } \left[ (1-x_t)^2 + 3(1-x_t y_t) \right] \ln(y_t)
  \right\}\,,  \nonumber \\
  C^{HH}_1 & = 2 x_t y_t (\zeta^u_{t q} \zeta^{u*}_{tt})^2 \left[ \frac{1+y_t}{2(1-y_t)^2} + \frac{y_t \ln(y_t)}{(1-y_t)^3}\right]\,, \nonumber \\
 C^{HH}_2 & =  4 (\zeta^u_{t q} \zeta^{d*}_{bb})^2 x_t y_t y_b \left[ \frac{2}{(1-y_t)^2} + \frac{(1+y_t) \ln(y_t) }{(1-y_t)^3}\right]\,.
 \end{align}
It is of interest to see that although $C^{HH}_2$ is proportional to the small factor $y_b$, due to $z^u_{23}\gg 1$ in this study, the $C^{HH}_2$ contribution is significant. If we take $y_b=\chi^u_{33,23}=0$,  our results are the same as those obtained in~\cite{Urban:1997gw}. Since we are interested in some what light charged-Higgs, i.e., $\mu_H$ is slightly higher than $m_t$, we take $\eta \approx 1$ in the numerical calculations. Thus, according to Eq.~(\ref{eq:mbWCs}) and the magic numbers in~\cite{Becirevic:2001jj}, the Wilson coefficients  $C_i(m_b)$ can be written as:
 \begin{align}
 C_1(m_b) \approx 0.848 C_1(\mu_H)\,, \ C_2(m_b) \approx 1.708 C_2(\mu_H)\,, \ C_3(m_b) \approx -0.016 C_2(\mu_H)\,.
 \end{align}
The matrix elements of the renormalized operators for $\Delta B=2$ are defined as~\cite{Becirevic:2001jj}:
 \begin{align}
 \langle B_{q} | \hat Q_1(\mu)  |  \bar B_{q} \rangle   &= \frac{1}{3} f^2_{B_{q}} m_{B_{q}} B_{1q}(\mu) \,, \nonumber \\
  \langle B_{q} | \hat Q_2(\mu)  | \bar B_{q}\rangle  &= -\frac{5}{24} \left(\frac{m_{B_{q}}}{m_b(\mu) + m_{q} (\mu)} \right)^2  f^2_{B_q} m_{B_q} B_{2q}(\mu) \,, \nonumber \\
   \langle B_{q} | \hat Q_3(\mu)  | \bar B_{q}\rangle  &= \frac{1}{24} \left(\frac{m_{B_{q}}}{m_b(\mu) + m_{q} (\mu)} \right)^2  f^2_{B_{q}} m_{B_{q}} B_{3q}(\mu) \,, 
 \end{align}
where the operators $\hat Q_{1,2,3}$, quark masses, and $B_{iq}$ parameters at $m_b$ scale  in the Landau RI-MOM scheme and the decay constants of $B_q$ are shown in Table~\ref{tab:BPs}~\cite{Becirevic:2001jj,Becirevic:2001xt,Becirevic:2001yv}. Due to $B_{is}\approx B_{id}$, we will adopt $B_{is}= B_{id}=B_{iq}$ in the numerical estimations. As a result,  $\langle B_{q}| H^{\rm \Delta B=2}_{\rm eff} | \bar B_{q} \rangle $ with the $W^\pm$- and $H^\pm$-boson contributions is given as:
 \begin{align}
\langle B_{q}| H^{\rm \Delta B=2}_{\rm eff} | \bar B_{q} \rangle&=  \langle B_{q}| H^{\rm \Delta B=2}_{\rm eff} | \bar B_{q} \rangle^{\rm SM} \left( 1 + \Delta^{H^\pm}_{q} \right)\,, \nonumber \\
\langle B_{q}| H^{\rm \Delta B=2}_{\rm eff} | \bar B_{q} \rangle ^{\rm SM} & = \frac{G^2_F (V^*_{tb} V_{tq})^2 }{48\pi^2} m^2_W f^2_{B_{q}}  m_{B_{q}} \hat{\eta}_{1B} B_{1q} ( 4S_0(x_t) )\,, \nonumber \\
 \Delta^{H^\pm}_q  = \frac{1}{4 S_0(x_t)} & \left[C^{WH}_1 + C^{HH}_1 +   \frac{ m^2_{B_{q}} C^{HH}_2  }{8 (m_b + m_{q})^2 \hat{\eta}_{1B} B_{1q} } \left( -5 \hat{\eta}_{2B} B_{2q} + \hat{\eta}_{3B} B_{3q} \right)  \right]
  \end{align}
where $\hat{\eta}_{1B}\approx 0.848$, $\hat{\eta}_{2B}\approx 1.78$, and $\hat{\eta}_{3B}\approx -0.016$ are the QCD corrections. The mass difference between the physical $B_q$ states can be obtained by:
 \begin{equation}
 \Delta M_q = 2 | \langle B_{q}| H^{\rm \Delta B=2}_{\rm eff} | \bar B_{q} \rangle|=\Delta M^{\rm SM}_q |1+\Delta^{H^\pm}_q|\,. 
 \label{eq:DMq}
 \end{equation}
 Taking $V_{td} \approx 0.0082 e^{- i \beta}$ with $\beta \approx 22.5^{\circ}$, $V_{ts} \approx -0.04$, and $m_t= \bar m_t (m_t) \approx 165$ GeV, the $B_q$-meson oscillation parameters $\Delta M_{d,s}$ in the SM are respectively obtained as:
 \begin{align}
 \Delta M^{\rm SM}_d &  \approx 3.32\times 10^{-13}\ \text{GeV} = 0.504\; {\rm ps}^{-1} \,, \nonumber \\
 \Delta M^{\rm SM}_s & \approx  1.16 \times 10^{-11}\ \text{GeV} = 17.60\; {\rm ps}^{-1}\,, 
 \end{align}
where the current data are $\Delta M^{\rm exp}_d = (0.5065 \pm 0.0019)$ ps$^{-1}$ and $\Delta M^{\rm exp}_s = (17.756 \pm 0.021)$ ps$^{-1}$~\cite{PDG}.  In order to consider the new physics contributions, when we use the $\Delta M^{\rm exp}_q$ to bound the free parameters, we take the SM predictions to be $\Delta M^{\rm SM}_d= 0.555^{+0.073}_{-0.046}$ ps$^{-1}$ and $\Delta M^{\rm SM}_s=16.8^{+2.6}_{-1.5}$ ps$^{-1}$~\cite{Lenz:2010gu}, in which the next-to-leading order (NLO) QCD corrections~\cite{Buras:1990fn,Ciuchini:1997bw,Buras:2000if} and the uncertainties from various parameters, such as CKM matrix elements, decay constants, and top-quark mass, are taken into account.  Hence, from Eq.~(\ref{eq:DMq}), the bounds from $\Delta B=2$ can be used as:
 \begin{align}
 0.76 \lesssim |1+ \Delta^{H^\pm}_d | \lesssim 1.15 \,, \nonumber \\
 0.87 \lesssim |1+ \Delta^{H^\pm}_s | \lesssim 1.38 \,. 
 \end{align}
\begin{table}[htp]
\caption{Values of quark masses and $B_{iq}$ parameters at $m_b$ scale in the RI-MOM scheme. The decay constants of the $B_{d,s}$ mesons are from~\cite{Lenz:2010gu}.}
\begin{tabular}{cccccccc}  \hline \hline
 $m_b$ & $m_s$ & $m_q$ & $B_{1q}$ & $B_{2q}$ & $B_{3q} $ & $f_{B_s}$ & $f_{B_d}$ \\ \hline 
  ~4.6 GeV~ & ~0.10 GeV~ & ~5.4 MeV~ & ~~0.87~~ & ~~0.82~~ & ~~1.02~~ & ~0.231 GeV ~& ~ 0.191 GeV\\ \hline \hline
\end{tabular}
\label{tab:BPs}
\end{table}%
\subsection{ Constraint from the $\bar B \to X_s \gamma$ process} 
In addition to the $\Delta B=2$ processes,  the penguin induced $b\to s \gamma$ decay is also sensitive to new physics. The current experimental value is $BR(\bar B \to X_s \gamma)^{\rm exp}=(3.32 \pm  0.15) \times 10^{-4}$ for $E_\gamma > 1.6$ GeV~\cite{Amhis:2016xyh}, and the  SM prediction with next-to-next-to-leading oder (NNLO) QCD corrections  is  $BR(\bar B \to X_s \gamma)^{\rm SM}=(3.36 \pm  0.23) \times 10^{-4}$~\cite{Czakon:2015exa,Misiak:2015xwa}.  Since the SM result is close to  the experimental data, $\bar B \to X_s \gamma$ will give a strict bound on the new physics effects. 

The effective Hamiltonian arisen from the $W^\pm$ and $H^\pm$ bosons for $b\to s \gamma$ at $\mu_b$ scale can be written as:
\begin{align}
{\cal H}_{b\to s \gamma} = -\frac{4G_F}{\sqrt{2}} V^*_{ts} V_{tb} \left(C_{7\gamma} (\mu_b)O_{7\gamma}  + C_{8\gamma} (\mu_b)Q_{8G} \right)\,, 
\end{align}
where the electromagnetic and gluonic dipole operators are given as:
 \begin{equation}
 O_{7\gamma} = \frac{e}{16\pi^2} m_b \bar s \sigma^{\mu\nu} P_R b F_{\mu\nu}\,, \  O_{8G} = \frac{g_s}{16\pi^2} m_b \bar s_\alpha \sigma^{\mu\nu}  T^a_{\alpha \beta} P_R b_\beta G^a_{\mu\nu}\,.
 \end{equation}
$C_{7\gamma}(\mu_b)$ and $C_{8G}(\mu_b)$ are the Wilson coefficients at $\mu_b$ scale, and their relations to the initial conditions at the high energy scale $\mu_H$ 
are through renormalization group (RG) equations. The NLO~\cite{Ciuchini:1997xe,Borzumati:1998tg,Borzumati:1998nx} and NNLO~\cite{Hermann:2012fc} QCD corrections to the $C_{7\gamma}(\mu_b)$ and $C_{8G}(\mu_b)$ in the 2HDM have been calculated.  Based on the $C^{\rm SM}_{7\gamma}(\mu_b)$ value extracted in~\cite{Blanke:2011ry}, we get  $C^{\rm SM}_{7\gamma}(\mu_b) \approx -0.304$ when $BR(\bar B \to X_s \gamma)^{\rm SM}=3.36 \times 10^{-4}$ is applied. 
In order to study the influence of the $b\to s \gamma$ process on the type-III model, we follow the approach  in~\cite{Misiak:2015xwa} and split the $BR(\bar B \to X_s \gamma)$ to be:
 \begin{align}
  BR(\bar B \to X_s \gamma) \times 10^{4} \approx (3.36 \pm 0.23) -8.22\, Re(C^{H^\pm}_{7\gamma}) -1.99\, Re(C^{H^\pm}_{8G})\,,
  \end{align}
 where $C^{H^\pm}_{7\gamma, 8G}$ are the Wilson coefficients at $\mu_H$ scale, (the matching scale is $\mu_0 \sim m_t$ at which the heavy particles are decoupled~\cite{Misiak:2015xwa}), and the quadratic $C^{H^\pm}_{7\gamma,8G}$  are ignored due to the requirement of $C^{H^\pm}_{7\gamma,8G}<1$.  Using the current experimental value, the bound on $C^{H^\pm}_{7\gamma,8G}$ is:
    \begin{equation}
 8.22 Re(C^{H^\pm}_{7\gamma}) + 1.99 Re(C^{H^\pm}_{8G}) \approx  0.04 \pm 0.28.  \label{eq:C7bound}
 \end{equation}
According to the charged-Higgs  interactions in Eq.~(\ref{eq:btoqpp}), the $H^\pm$ contributions to $C^{H^\pm}_{7\gamma,8G}$ are expressed as~\cite{Borzumati:1998tg}: 
 \begin{align}
 C^{H^\pm}_{7\gamma} & = \zeta^u_{tt} \zeta^{u*}_{ts} C^{H^\pm}_{7,LL} +  \zeta^d_{bb} \zeta^{u*}_{ts} C^{H^\pm}_{7,RL}  \,, \nonumber \\
C^{H^\pm}_{8G} &= \zeta^u_{tt} \zeta^{u*}_{ts} C^{H^\pm}_{8,LL} +  \zeta^d_{bb} \zeta^{u*}_{ts} C^{H^\pm}_{8,RL} \,, \label{eq:CCH7}
 \end{align}
 \begin{align}
 C^{H^\pm}_{7,LL} & = \frac{y_t}{72} \left[ \frac{8 y^2_t + 5 y_t -7}{(1-y_t)^3} - \frac{6 y_t (2-3y_t)}{(1-y_t)^4} \ln(y_t)\right]\,, \nonumber \\
 C^{H^\pm}_{8,LL} & = \frac{y_t}{24} \left[ \frac{ y^2_t - 5 y_t -2}{(1-y_t)^3} - \frac{6 y_t }{(1-y_t)^4} \ln(y_t)\right]\,, \nonumber \\
 C^{H^\pm}_{7,RL} & = \frac{y_t}{12}  \left[ \frac{3-5y_t}{(1-y_t)^2} + \frac{2(2-3y_t)}{(1-y_t)^3 }\ln(y_t)  \right]\,, \nonumber \\
 C^{H^\pm}_{8,RL} & = \frac{y_t}{4}  \left[ \frac{3-y_t}{(1-y_t)^2} + \frac{2}{(1-y_t)^3 } \ln(y_t)\right]\,.
 \end{align}
 Taking $\chi^u_{33,23}=\chi^d_{33}=0$ in Eq.~(\ref{eq:CCH7}), it can be found that $\zeta^u_{tt} \zeta^{u*}_{ts}$ in type-II 2HDM  is suppressed by $1/t^2_b$ while the $t_\beta$-dependence  in $\zeta^{d}_{bb}$ $\zeta^{u*}_{ts}$ is canceled and $(\zeta^d_{bb}$ $\zeta^{u*}_{ts})_{\rm type-II}=1$. As a result,  the mass of charged-Higgs in type-II 2HDM has been limited to be $m_{H^\pm} > 580$ GeV at 95\% confidence level (CL) by using NNLO QCD corrections~\cite{Misiak:2017bgg}. 

\begin{figure}[phtb]
\includegraphics[scale=0.4]{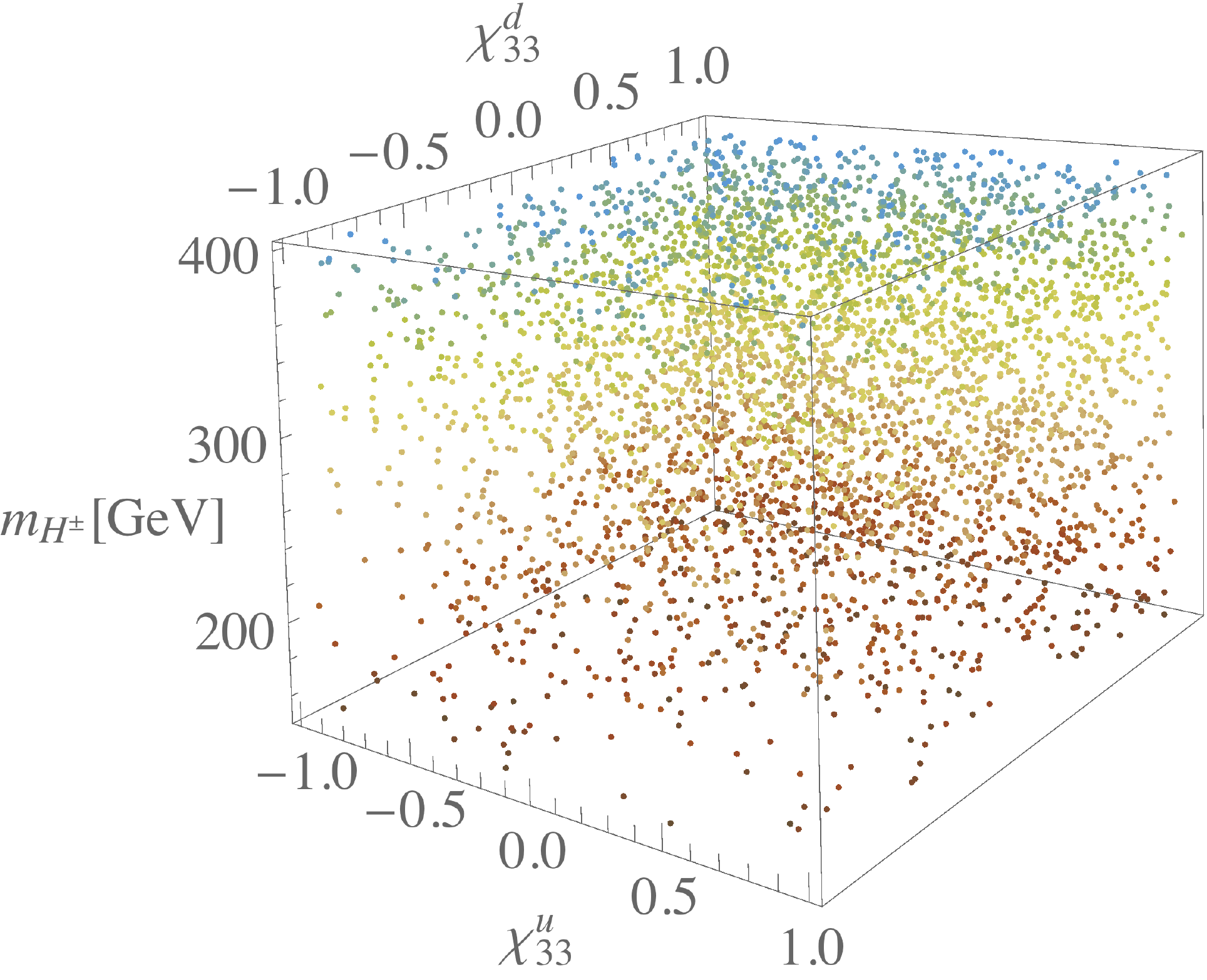}
\includegraphics[scale=0.3]{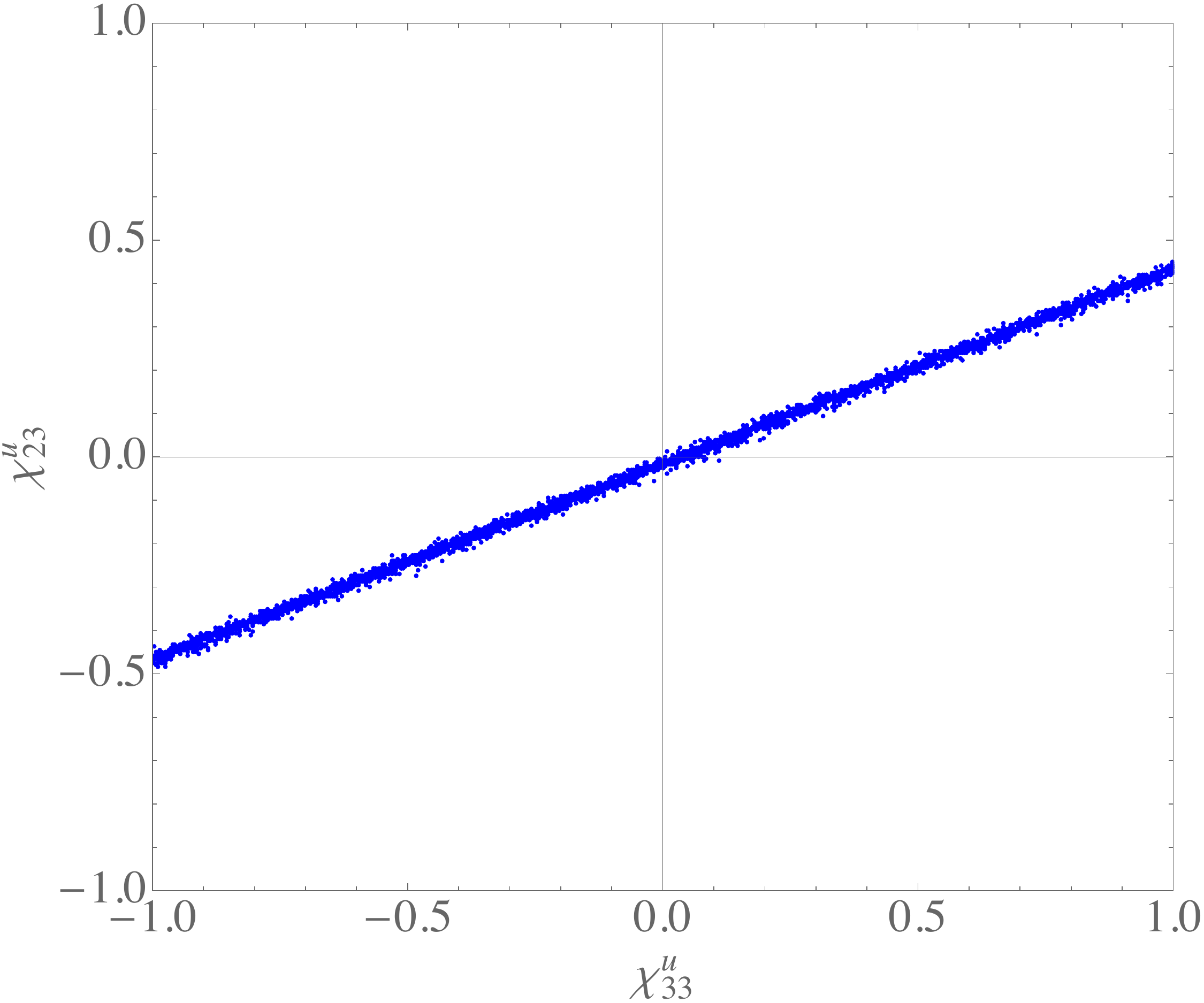}
\caption{ Left panel: allowed  parameter spaces of $(m_{H^\pm}, \chi^u_{33}, \chi^d_{33})$ with $t_\beta=[30,\; 60]$ when  the bound from $\bar B\to X_s \gamma$ shown in Eq.~(\ref{eq:C7bound}) is satisfied. Right panel:  the results of left panel project onto $\chi^u_{33}$-$\chi^u_{23}$ plane, where   
$t_\beta=[30,\; 60]$, $m_{H^\pm}=[150,\, 400]$ 
GeV, and $\chi^u_{33,23}, \chi^d_{33}=[-3,\; 3]$  are included. }
\label{fig:3dplot}
\end{figure}

In the type-III 2HDM, it is observed that in the large $tan\beta$ region, due
to the $1/c_\beta$ enhancement, the $\zeta^d_{bb} \zeta^{u*}_{ts}$ terms still
dominate. Since the new parameters $\chi^u_{33,23}/c_\beta$ and
$\chi^d_{33}/s_\beta$ are involved in Eq.~(\ref{eq:CCH7}), it is possible to reduce ($\zeta^d_{bb}$ $\zeta^{u*}_{ts})_{\rm type-II}$ far away from unity; thus, the charged-Higgs mass can be much lighter than 580 GeV which can be seen in Figure.(\ref{fig:3dplot}) left panel. In the other hand, we can get constraints on $\chi_{33}^u$ and $\chi_{23}^u$ from the right panel of the same Figure.(\ref{fig:3dplot}), it is clear that for $-1\leqslant \chi^u_{33} \leqslant 0 $ we obtain $-0.5 \leqslant \chi^u_{23} \leqslant 0 $ and for $0 \leqslant \chi^u_{33} \leqslant 1 $ the interval permitted for $\chi^u_{23}$ become $[0,\; 0.5]$.
 
\section{Constraints from $B_{q}\to \mu^+ \mu^-$ and $B_{q}\to X_s \mu^+ \mu^-$}
\label{sec:BsmmandBsmm}
\subsection{Constraints from $B_{q}\to \mu^+ \mu^-$, ($q=s,d$) }
\begin{figure}[phtb]
\includegraphics[scale=0.83]{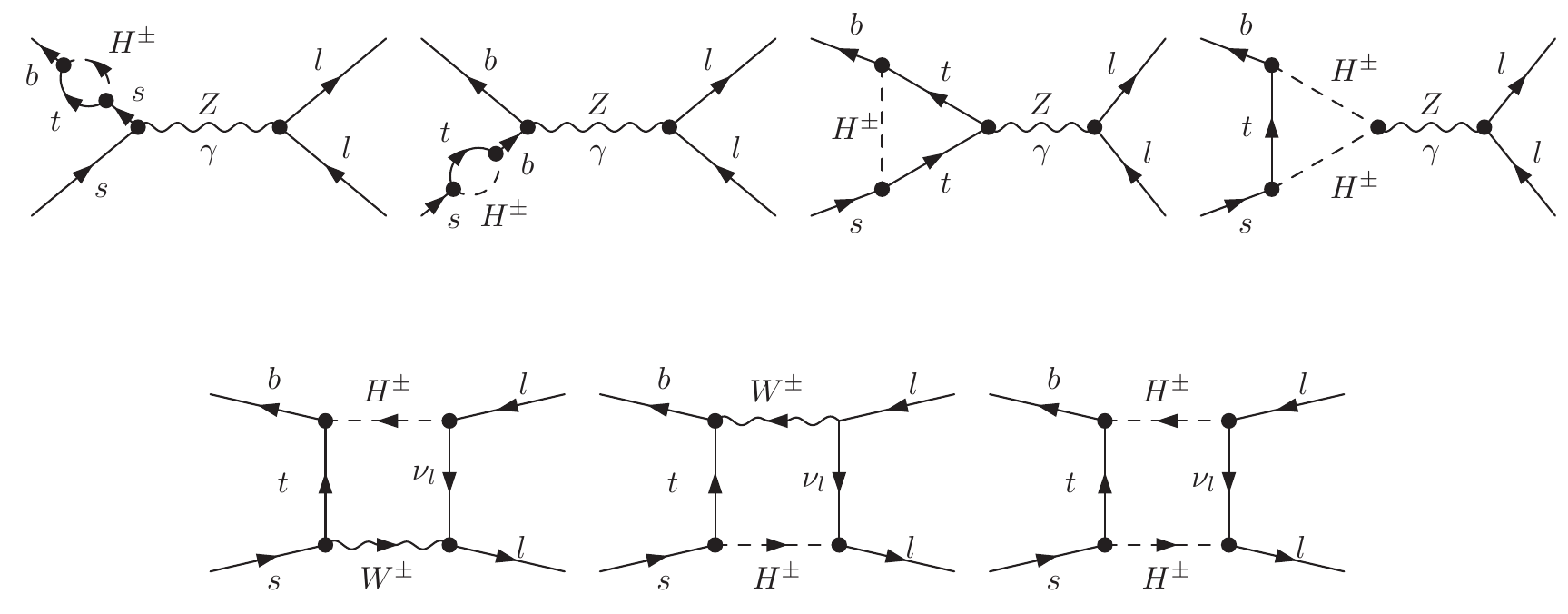}
\caption{ The sketched electroweak penguin and box diagrams for $b\to s \ell^+ \ell^-$ 
mediated by $H^\pm$ and $W^\pm$ bosons. }
\label{fig:HpC}
\end{figure}

The effective Hamiltonian for $\Delta B = 1$ can be written as \cite{Chetyrkin:1996vx,Buchalla:1995vs, Altmannshofer:2008dz}:
\begin{align}
{\cal H}_{eff} = -\frac{G_F}{\sqrt{2}}\alpha 
\left( V^*_{ts} V_{tb}\sum_{i}\left( {\cal C}_{i} {\cal O}_{i} + 
{\cal C}^{\prime}_{i} {\cal O}^{\prime}_{i}\right) . h.c\right)\,, 
\end{align}
Where $ {\cal C}_{i} $ and ${\cal C}^{\prime}_{i}$ are Wilson coefficients encoding the short-distance physics at the energy scale $\mu$ which is usually taken to be the $b$-quark mass ($m_b$), and can be modified from SM predictions in the presence on the new physics, while ${\cal O}_{i}$ are the operators given by 
\begin{eqnarray}
{\cal O}_{9} &=&  \left(\bar{s} \gamma_\mu P_L b \right) \left(\bar{\ell} \gamma^\mu \ell\right),\,\,\,
{\cal O}_{10} = \left(\bar{s}\gamma_\mu P_L b \right) \left(\bar{\ell}\gamma^\mu \gamma_5  \ell\right),\,\, \\
{\cal O}_{S} &=&  \left(\bar{s} P_R b \right) \left(\bar{\ell}  \ell\right),\,\,\,{\cal O}_{P} =  \left(\bar{s} P_R b \right) \left(\bar{\ell} \gamma_5 \ell\right),\,\,\,
 \label{eq:operators}
\end{eqnarray}
with $P_{L,R} = (1\mp \gamma_5)/2$ and $m_b = m_b(\mu_b)$ denotes the running $b$ quark mass in the $\overline{MS}$ scheme with $\mu_b = 4.8$ GeV. The ${\cal O}^{\prime}_{9,10}$ can be obtained from the ${\cal O}_{i}$ by making the replacements $P_L \leftrightarrow P_R$. In the SM, three operators play an important role, namely the electromagnetic operator ${\cal O}_{7}$, and the semileptonic operators ${\cal O}_{9,10}$, differing with respect to the chirality of the emitted charged leptons \cite{DescotesGenon:2011yn}. 
The SM values ${\cal C}_{9}$ and ${\cal C}_{9}$ are obtained at the next-to-next-to-leading order (NNLO)\cite{Bobeth:2003at, Huber:2005ig} and depend on the fundamental parameters of the top-quark mass and $W$-boson masses as well as the weak mixing angle $\theta_W$. Moreover, they are universal for the three lepton flavors $\ell = e, \mu, \tau$. The other Wilson coefficients ${\cal C}^{\prime}_{i}$ are suppressed by $m_b m_{\ell}/m^2_W$. 
The Wilson coefficient in the SM are $ {\cal C}_{9} = 4.211$ and ${\cal C}_{10} = {\cal C}^{SM}_{10}$ = - $\eta_Y Y_0 (m^2_t/m^2_W)/\sin^2\theta_W = -4.103$ \cite{Huber:2005ig,Bobeth:1999mk}, where $Y_0$ is one-loop function~\cite{Inami:1980fz} and $\eta_Y = 1.026 \pm 0.006$ summarizes the NLO corrections~\cite{Inami:1980fz} with $m_t = \bar{m_t} (m_t)$. 
In the general 2HDM, $b \to s$ transition is mediated by gauge boson $Z$, Goldsotne boson $G^0$, and neural Higgs bosons h, H and A penguin diagrams, as well as box diagrams mediated with $W^\pm$, $H^\pm$ and $G^\pm$ which lead  to additional contributions 
to ${\cal C}_{i}$ ($i=7,9,10$) and could make the chirality-flipped operators ${\cal O}_{i}$ ($i=7,9,10$) to contribute in a significant manner through the $Z$- and $\gamma$ diagrams shown in Fig.\ref{fig:HpC}. 
In the following we separate the contribution in two categories penguins and boxes, the scalar charged Higgs boson contribution comes from Z and $\gamma$-penguin are :
\begin{eqnarray}
C_9 &\nonumber=& \xi^{u}_{tt}\xi^{u*}_{ts} f_{3}(x_t,y_{H^\pm}) + \xi^{d}_{bb}\xi^{u*}_{ts} x_d f_{4}(x_t,y_{H^\pm}) + \eta_W C_{10} - 
4 \eta^\prime_W { \rm Re}(\xi^{u}_{tt}\xi^{\mu*}_{22} x_\mu f_{5}(x_t,y_{H^\pm})) \\&+& 
\eta^\prime_W \xi^{u}_{tt}\xi^{u*}_{ts} |\xi^{\mu}_{22}|^2  x_\mu f_{6}(x_t,y_{H^\pm}) + \eta^\prime_W { \rm Re} (\xi^{u}_{tt}\xi^{\mu*}_{22}) x_\mu f_{7}(x_t,y_{H^\pm})\\
C_{10} &=& \eta^\prime_W \bigg(\xi^{u}_{tt}\xi^{u*}_{ts} f_{8}(x_t,y_{H^\pm}) + \xi^{d}_{bb}\xi^{u*}_{ts} x_b f_{9}(x_t,y_{H^\pm}) + \xi^{u}_{tt}\xi^{u*}_{ts} |\xi^{\mu}_{22}|^2 x_\mu f_{10}(x_t,y_{H^\pm}) \\\nonumber&+&  {\rm Re} (\xi^{u}_{tt}\xi^{\mu*}_{22}) x_\mu f_{11}(x_t,y_{H^\pm})\bigg)\\
C_P &=& \sqrt{x_b x_\mu} \bigg[\xi^{u*}_{tt}\xi^d_{bb}\eta^\prime_W g_{1}(x_t,y_{H^\pm}) + \xi^{u}_{tt}\xi^{u*}_{ts} \bigg( g_{2}(x_t,y_{H^\pm}) - \eta^\prime_W g_{3}(x_t,y_{H^\pm})\bigg) 
\\\nonumber &+&
\eta^\prime_W \bigg(\xi^{\mu}_{22}\xi^{u*}_{tt} g_{4}(x_t,y_{H^\pm}) - \xi^{\mu *}_{22}\xi^{u}_{tt} g_{5}(x_t,y_{H^\pm}) - 2\xi^{d}_{bb}\xi^{\mu *}_{22}  g_{6}(x_t,y_{H^\pm})\bigg)\bigg] \\
C_S &=& \sqrt{x_b x_\mu} \eta^\prime_W \bigg[ \xi^{\mu}_{22}\xi^{u*}_{tt} g_{4}(x_t,y_{H^\pm}) + \xi^{\mu *}_{22}\xi^{u}_{tt} g_{5}(x_t,y_{H^\pm}) + 2\xi^{d}_{bb}\xi^{\mu*}_{22} g_{6}(x_t,y_{H^\pm})\bigg]
\end{eqnarray}
where $\eta_W = (-1 + 4\sin^2\theta_W)$ and $\eta^\prime_W = \sin^{-2}\theta_W$. With $ x_i = m^2_i/m^2_W $ with $i= t,b, \mu$ 
and $y_{H^\pm} = m^2_{H^\pm}/m^2_W $. The corresponding loop functions $f_i$ and $g_i$ can be found in Appendix A. 
In what follow, we will concentrate our discussion on the Wilson coefficients $C_9$ and $C_{10}$ which can be extracted from different  angular observables, in particular in the case of $B \to K^* \mu^+ \mu^-$ which provides a several observables through angular study of the decay which have been experimentally studied at LHCb\cite{Aaij:2013iag, Aaij:2015oid}
, CMS\cite{Chatrchyan:2013cda,Khachatryan:2015isa}, ATLAS\cite{Aaltonen:2011ja}, Belle\cite{Wei:2009zv, Wehle:2016yoi} and BABAR \cite{Aubert:2006vb}.
Several observables have shown deviations from SM predictions. It started with the set of observables $P^\prime_5$, $Q_5 = P^{\prime\mu}_5 - P^{\prime e}_5$, forward-backward asymmetry ($A_{FB}$), 
lepton-flavour universality violating ratio $R_{K^*}$. 

\begin{figure}[phtb]
\includegraphics[scale=0.55]{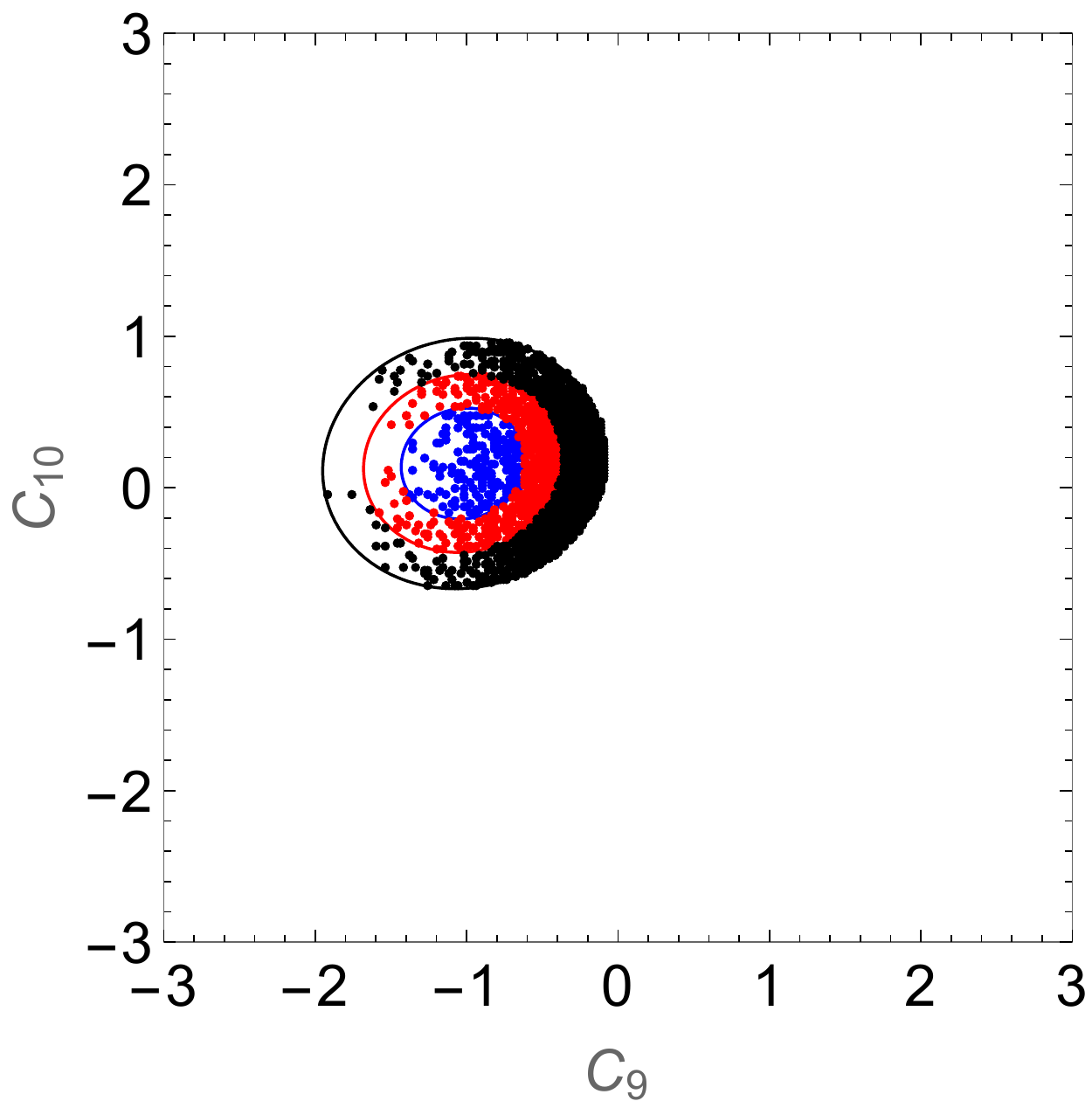}
\includegraphics[scale=0.57]{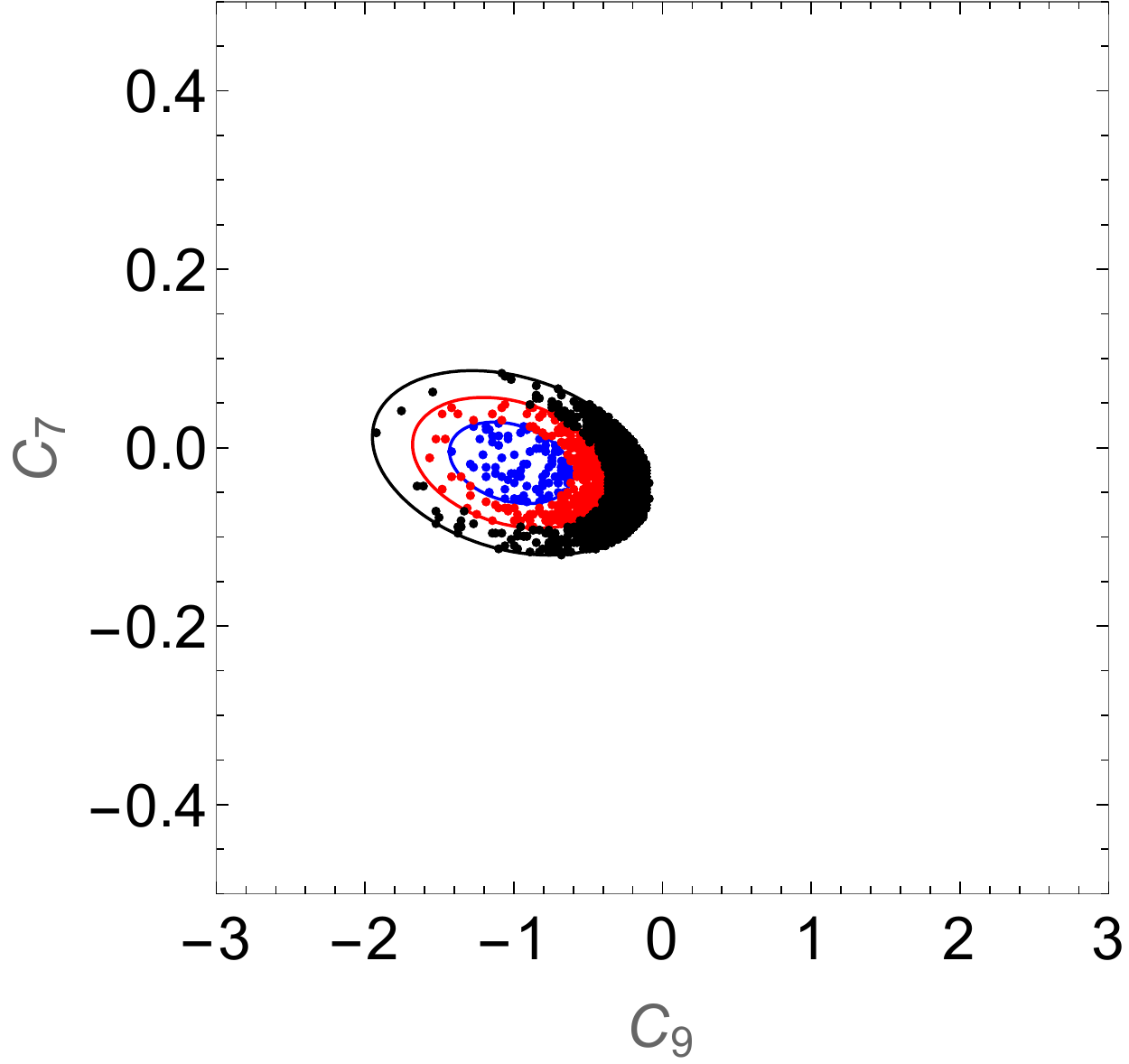}
\caption{ Correlation between Wilson coefficients in the allowed region of $BR(b \to s \gamma)$ at 95$\%$ CL and the Eq.(\ref{eq:fits}).}
\label{fig4:c7910}
\end{figure}

Several global fits exist for NP contributions to the Wilson coefficients $C_{9, 10}$ \cite{Altmannshofer:2014rta, Descotes-Genon:2015uva, Descotes-Genon:2013wba}. These fits includes the branching ratios of $B\to K \mu^+ \mu^-$, $B\to K^* \mu^+ \mu^-$, $B_s \to \phi\mu^+\mu-$, $B_s\to X_s \mu^+ \mu^-$ (restricted only to the range $q^2\in$ [1,6] GeV$^2$), $B \to X_s \gamma$, $B_s \to \mu^+\mu^-$ as well as some isospin symmetry and time-dependent CP asymmetry of $B\to K^* \gamma$. To be more conservative, we use the central values given by:
\begin{eqnarray}
 C_7 = -0.017 \pm 0.030, \,\,\, C_9 = -1.02 \pm 0.27, \,\,\, C_{10} = 0.16 \pm 0.24, \,\,\,
\label{eq:fits}
\end{eqnarray}
and we use the following correlation coefficients $\rho_{C_7, C_9} = -0.28$
and  $\rho_{C_9, C_{10}} = +0.06$. These bounds can be used to impose
constraints on our parameters space. We show in Figure.(\ref{fig4:c7910}) the
correlation between Wilson coefficients in the allowed region with the same
parameters as in Figure.(\ref{fig:3dplot}) and taking into account constraints from $BR(b \to s \gamma)$ at 95$\%$ CL. As can be seen $C_9 < 0$ is preferred by data,  and the possibility $C_9 = - C_{10}$ can also be a good fit. 
\begin{table}[htp]
\caption{Current measurement with 1 GeV$^2$ $< q^2 < 6$ GeV$^2$.}
\begin{center}
\begin{tabular}{cc}  \hline \hline
 BR$(B\to X_s \mu^+ \mu^-)$\cite{Lees:2013nxa} &  BR$(B_s \to  \mu^+ \mu^-)$\cite{Patrignani:2016xqp}\\ \hline 
  ~$(0.66^{+0.82 + 0.30}_{-0.76 - 0.24} \pm 0.07)\times 10^{-6}$~ &  $(2.4^{+0.9}_{-0.7})\times 10^{-9}$ \\ \hline \hline
\end{tabular}
\end{center}
\label{tab:BRPs}
\end{table}%
The expression for the branching ratio $B_s \to \mu^+ \mu^-$ is given by
\begin{eqnarray}
BR(B_s \to \mu^+ \mu^-) &=& \tau_{B_s} m^3_{B_s} f^2_{B_S} m^2_\mu \beta\frac{\alpha^2 G^2_F}{16\pi^3} |V_{tb}V_{ts}|^2
\bigg[\bigg|C_{10} + \frac{m^2_{B_s}}{2m_\ell(m_b + m_s)} C_P\bigg|^2 \\\nonumber &+& 
\frac{m^4_{B_s} \beta^2}{4m^2_{\ell}(m_b + m_s)^2}\bigg| C_S \bigg|^2\bigg]
\end{eqnarray}
with $\beta = (1-4m^2_\mu/m^2_{B_s})^{1/2}$, where $C_9$ does not contribute.  In the above relation BR$(B_s \to \mu^+ \mu^-)$ is calculated by ignoring the 
 $B_s - \bar B_s$ mixing. Experiments measure the average time-integrated 
branching ratio denoted by $\overline{BR} (B_s \to \mu^+ \mu^-)$ and the two are related as
$\overline{BR}(B_s \to \mu^+ \mu^-) = \frac{1+y_s \Delta A_s}{1-y^2_s} BR(B_s
\to \mu^+ \mu^-)$. 
Where $y_s = 0.062\pm 0.006$ \cite{Beaujean:2013soa, Amhis:2016xyh, PDG} and $\Delta A_s$ is the CP asymmetry due to vanishing width difference, which is $\Delta A_s  = +1$ in the SM, but in general it can be $\Delta A_s\in[-1,1]$ \cite{Fleischer:2017yox}. The SM prediction of the branching ratio of a $B_s$ meson decaying into two muons is calculated to be
$\overline{BR}(B_s \to \mu^+ \mu^-)_{SM} = (3.66 \pm 0.23)\times 10^{-9}$.
This prediction is particularly precise thanks to the purely leptonic final state which reduce the dependence on computations of the strong force. 
A fit to the invariant mass of the dimuon candidates $m_{\mu\mu}$ have been 
performed by
LHCb and CMS groups. The measured Branching ratio is given in Table.\ref{tab:BRPs}
with 6.2 standard deviations. 
In order to probe new physics effects, it's convenient to define the quantity $R_{B_s}$ by
\begin{eqnarray}
R_{B_s} = \frac{\overline{BR}(B_s \to \mu^+ \mu^-)}{\overline{BR}(B_s \to \mu^+ \mu^-)_{SM}} = \big|P\big|^2 + \big|S\big|^2
\end{eqnarray} 
where $S$ and $P$ are given by\cite{Fleischer:2017ltw}
\begin{eqnarray} 
P = \frac{C_{10}}{C^{SM}_{10}}  + \frac{m^2_{B_s}  }{2 m_\mu(m_b + m_s)} \frac{C_P}{C^{SM}_{10}} \,\,\,\quad {\rm and} \quad\,\,\,
S = \frac{m^2_{B_s}  \beta }{ 2 m_{\mu}(m_b + m_s)} \frac{C_S}{C^{SM}_{10}}
\label{eq:SP}
\end{eqnarray} 
In the SM, $C_{10} = C^{SM}_{10}$, then $P_{SM} = 1$ and $S_{SM} = 0.$ Combining the experimental values we get:  
$R_{B_s} = 0.90^{+ 0.42}_{-0.34}$ . Although the uncertainty in this ratio is somewhat large, values smaller than unity seem to be preferred.
In type-II 2HDM, the $C_S$ and $C_P$ contributions to Eq.(\ref{eq:SP}) are 
suppressed \cite{Li:2014jsa} unless for large Yukawa couplings, however in type-III 2HDM, it gets large enhancements from the $ \xi^{u,d}_{tt, bb}$ parameters, where the contribution depends also on $m_{H^\pm}$ and $\tan\beta$.

\begin{figure}[phtb]
\begin{center}
\includegraphics[scale=0.57]{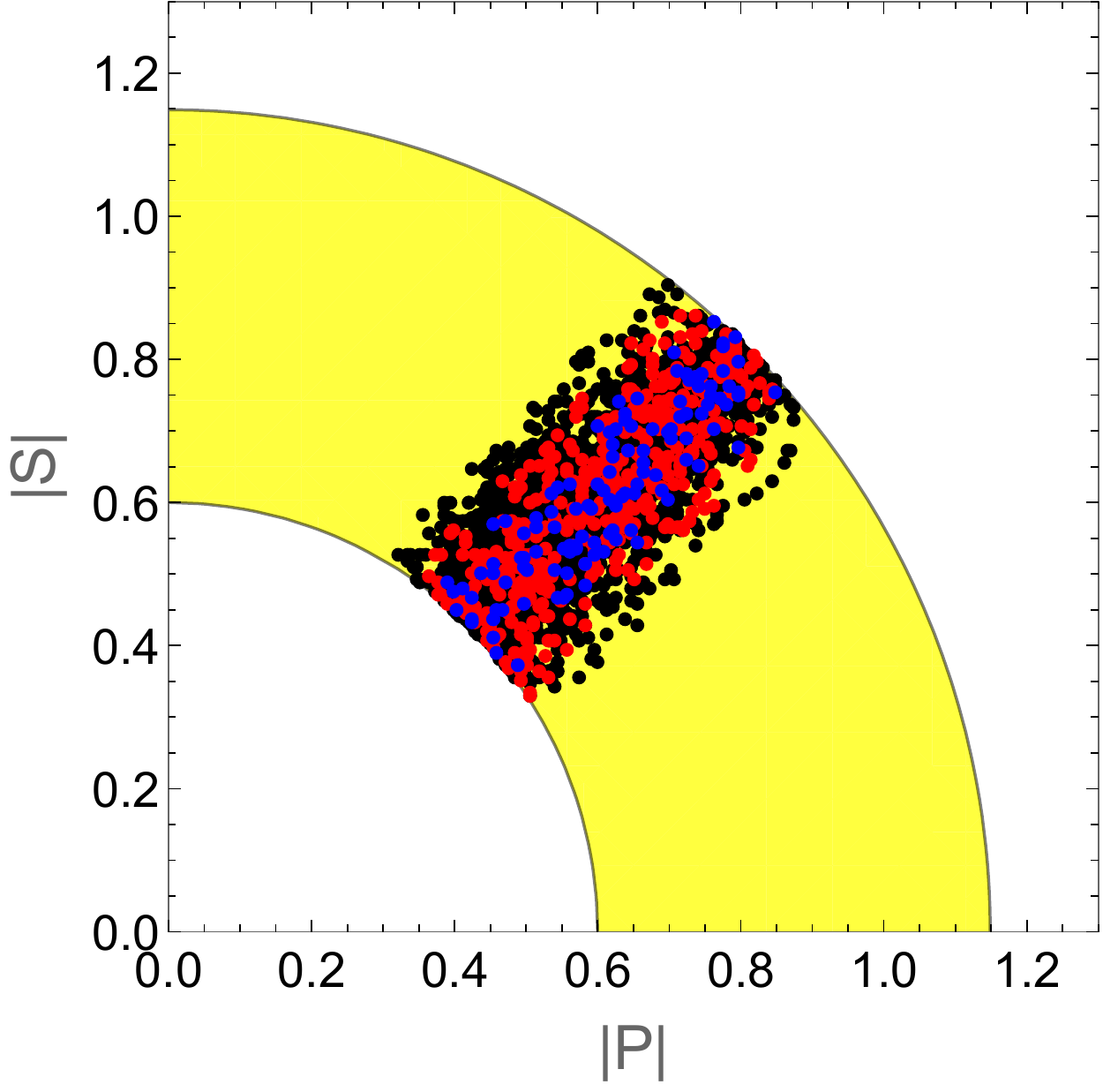}
\end{center}
\caption{Correlations between Wilson coefficients $P$ and $S$. Taking into 
account constraints from Eq.~(\ref{eq:fits}) as well as from 
$BR(b\to s \gamma)$.
Yellow band represent 2$\sigma$ of $R_{B_s}$.}
\label{fig4:3dplot}
\end{figure} 
Other complementary information on Wislon coefficients can be extracted from
the decay $B_q \to X_s \mu^+ \mu^-$ branching ratios in the range 1 GeV$^2$ $<
q^2 = m^2_{\mu\mu} <$ 6  GeV$^2$. 
We use the integrated rate as given in Ref \cite{Huber:2005ig}:
\begin{eqnarray}
BR(B_q\to X_s \mu^+ \mu^-) \nonumber &=& \bigg( 2.1913 - 0.001655 {\cal I}(r_{10}) + 0.0535{\cal I}(r_{7}) + 0.00496{\cal I}(r_{7} r^*_9) \\\nonumber &-& 0.0118{\cal I}(r_{9})  - 0.5426{\cal R}(r_{10}) + 0.0281 {\cal R}(r_{7}) + 0.0153{\cal I}(r^*_{10}r_7) \\\nonumber &-&0.8554 {\cal I}(r_{7} r^*_9) + 2.7008 {\cal R}(r_{9})  - 0.10705 {\cal I}(r_9 r^*_{10}) + 10.7687 |r_{10}|^2  \\ &+& 0.2889 |r_{7}|^2 + 1.4882 |r_{9}|^2
\bigg)\times 10^{-7}
\end{eqnarray}
where $r_i = C_{i}/C^{SM}_{i} $. The SM predictions is 
$ BR(B_q \to X_s \mu^+ \mu^-) = (1.59 \pm 0.11)\times 10^{-6}$.

\begin{figure}[phtb]
\hspace{-2.55cm}
\includegraphics[scale=0.9]{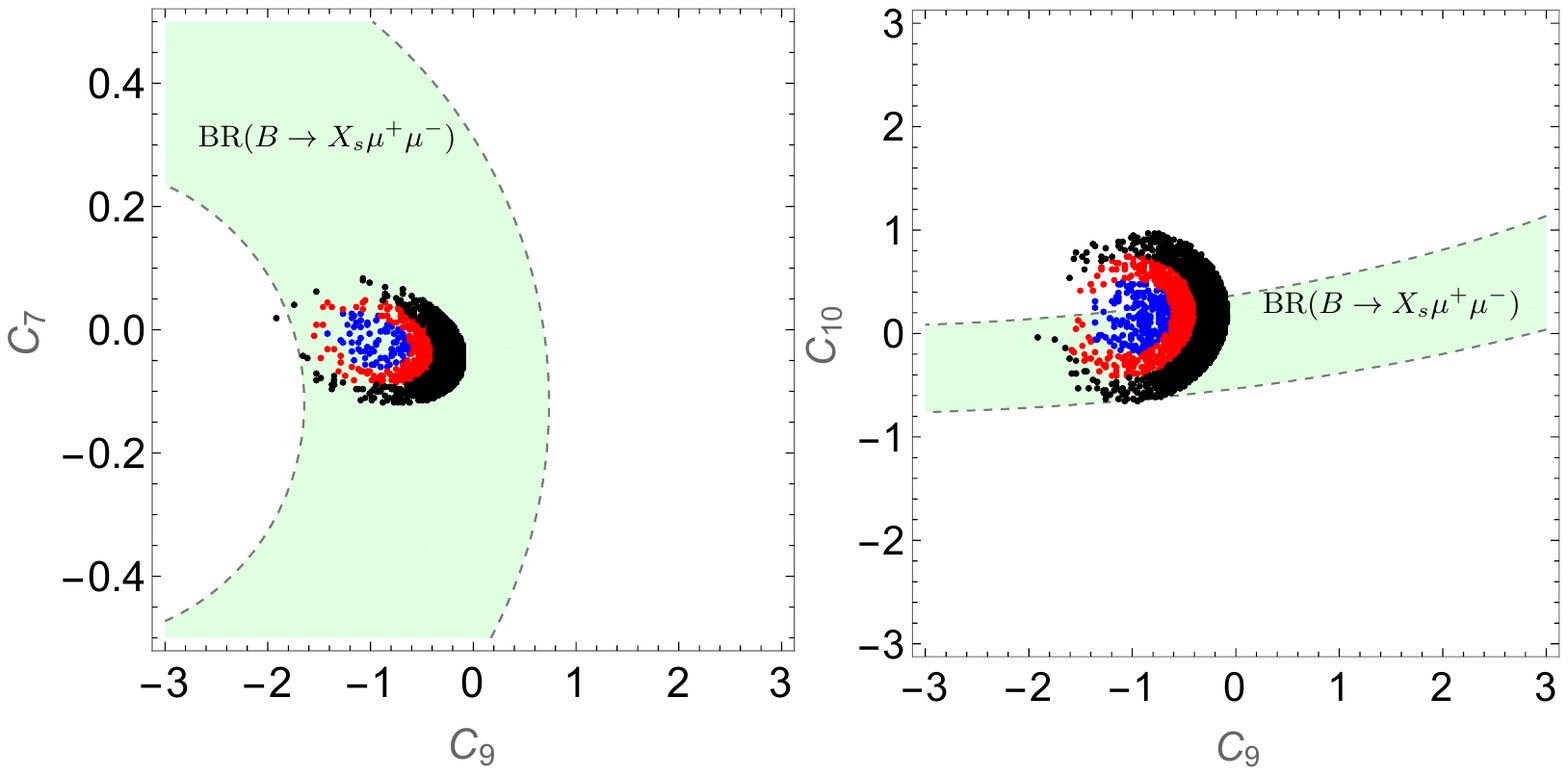}
\vspace{-18.1cm}
\caption{90$\%$CL bounds in the ($C_9, C_{10}$)(left) and ($C_9, C_{7}$)(right) planes following the experimental branching ratio of $ BR(B_q \to X_s \mu^+ \mu^-)$. 
The scatter points correspond to expectation in type-III-2HDM.}
\label{fig4:c7910Bsmm}
\end{figure} 
We impose the experimental bound on $BR(B \to X_s \mu^+ \mu^-)$ at 90$\%$CL and include constraints from Eq.(\ref{eq:fits}). We present in Fig.\ref{fig4:c7910Bsmm} the resulting allowed scatter points in the ($C_9, C_{10}$)(left) plane with $C_7 = -0.017$ and $C_{10} = 0.16$ in the right plot. The regions in Fig.\ref{fig4:c7910Bsmm} suggest that the best fit to data is achieved if non-zero contributions are present for $C_{7,9,10}$ Wilson coefficient that involve muons and those $C_{7,9,10}\neq 0$ seem to be preferred.
\section{Predictions of $R_K$ and $R_{K^*}$ in type III of 2HDM}
\label{sec:RKRKs}
In terms of the operators of the type (\ref{eq:operators}).The dependence on the Wilson coefficients of $R_K$ and $R_{K^*}$ in the bins $[1,6]$ GeV$^2$ and $[1.1,6]$ GeV$^2$, respectively can be expressed as \cite{Jager:2014rwa}:
\begin{eqnarray}
R_K &=& 10^{-2}\bigg( 2.9438 \left( |C_9|^2 + |C_{10}|^2\right) - 2{\rm Re}(C_9(0.8152 + i 0.0892)) + 0.2298\bigg),\\
R^{low}_{K^*} &=& 10^{-2}\bigg( 3.586 \left( |C_9|^2 + |C_{10}|^2\right) - 2{\rm Re}(C_9(2.021 + i 0.188)) \\\nonumber  &-& 2{\rm Re}(C_9(5.255 + i 0.239)) + 31.658 \bigg).
\label{eq:RK-RKs}
\end{eqnarray}
\begin{figure}[phtb]
\hspace{-2.54cm}
\includegraphics[scale=0.90]{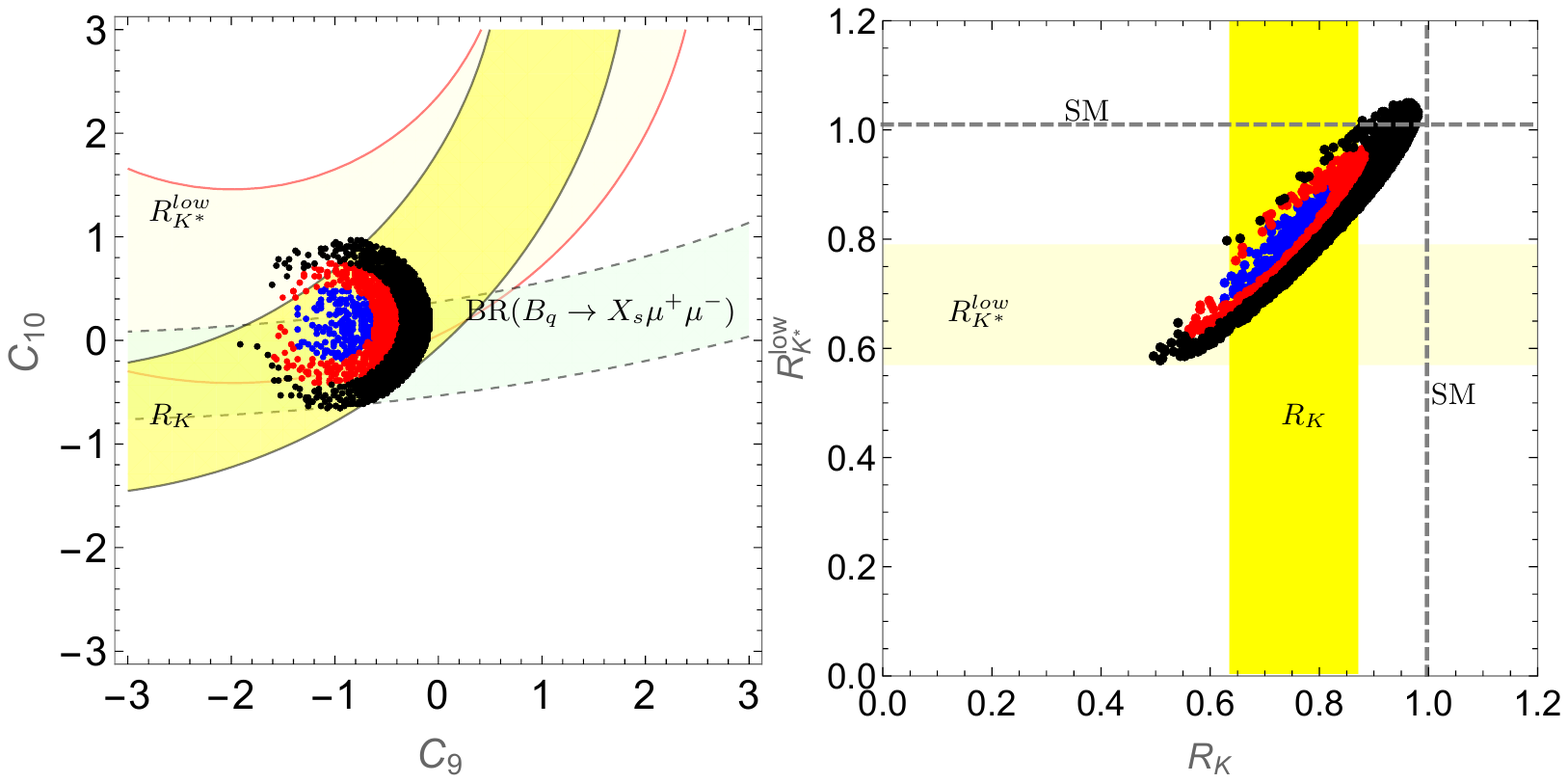}
\vspace{-18.0cm}
\caption{Right: Correlations between $R_K$ and $R^{low}_{K^*}$ in type-III 2HDM by taking all in the text. Left: projection of all constraints in the ($C_9$, $C_{10}$) plane.}
\label{fig4:3dplot}
\end{figure} 
It is clear that a decrease in $R_{K^{(*)}}$ compared to the SM prediction can 
 be achieved only for $C_{9,10} = 0$. The NP contributions to the Wislon
 coefficients have further consequences than simply altering the $R_{K^{(*)}}$
 observables, and it is crucial to notice that the sizes of $C_{9,10}$ that
 are allowed in type-III 2HDM in order  to accommodate the $R_{K^{(*)}}$ anomalies are also in the region preferred by $b\to s $ transitions, for instance, $B_q \to X_s \mu^+ \mu^-$, $B_q \to \mu^+ \mu^-$ and $B\to X_s \gamma$. In Fig.\ref{fig4:3dplot}(left), the projected constraints from $R_K$ and $R^{low}_{K^*}$ and together with $B \to X_s \mu^+ \mu^-$ are shown within the current experimental limit (1$\sigma$). The values of $\chi^{u,d}_{ij}$, $\tan\beta$ and $m_{H^\pm}$ varied as in Fig.\ref{fig:3dplot}, compatible regions where obtained for large $|\chi^{u,d}_{33}|, |\chi^{\mu}_{22}|\sim 3$ and positive $\chi^{d}_{23}\sim 1$. 
Given that $|C_9| \sim |C_{10}|$ is the most preferred scenario, it becomes obvious that $R_K$ $\sim$ $R^{low}_{K^*}$ as shown in the right plot of Fig.\ref{fig4:3dplot} with the current experimental limit (1$\sigma$), and the SM lines are also shown. Interestingly, accommodating the $R_{K^{(*)}}$ implies that the value of $R_{B_s}$ should deviate from the SM prediction by 10$\%$ from the central value. A more precise measurement of $B_s \to \mu^+ \mu^-$ branching fraction will provide more information on $C_{10}$ and so that $R_{K^{(*)}}$.

\section{Summary}
\label{sec:conlusion}
The recent years of activity at the LHC have brought to light several anomalies in exclusive semileptonic decays. Even though the latest model independent analyses are pointing to sizable NP contributions to different Wilson coefficients. The possibility of interpreting these results in the current situation of the SM is not possible, so it would be worth studying the possibility beyond the SM such as general 2HDM. 

In this work, we have studied these anomalies in the context of type-III 2HDM, unlike 2HDM-II, 
it would be still possible to have relatively light charged scalar in the
range $200$-$400$ GeV. By taking constraints from $\Delta M_q$ ($q=s,d$), $B \to  X_s \gamma$  $B_s \to \mu^+ \mu^{-}$ and $B_q \to X_s \mu^+ \mu^{-}$  we have studied the implications of G2HDM on $ R_{K^{(*)}}$ to identify how large deviations from the SM predictions are possible.

To obtain compatible $ R_{K^{(*)}}$ measured by LHCb, Belle and BaBar a
scenario with a large negative $C_9$ is found due to the charged-scalar
exchanges through the Z- and $\gamma$- penguin diagrams and under the
assumptions of light charged Higgs, large $\tan\beta$ with moderate Yukawa
couplings of the order of ${\cal O}(1)$. Moreover, of particular interest is the $R_{B_s}$ ratio whose value has been measured to be smaller than its SM prediction by a factor of 10$\%$. In type-III 2HDM we found that $ R_{K^{(*)}}$ are predicted to be similar.
Finally, future precise measurement of the $ R_{K^{(*)}}$ would be very helpful to provide a more definite answer concerning $b\to s$ transitions at the LHCb, Belle and BaBar collaborations restricting further or even deciphering the NP models.

\section*{Acknowledgments}
RB was supported in part by Chinese Academy of Sciences (CAS) President's International Fellowship Initiative (PIFI) program (Grant No. 2017VMB0021).  This work is supported also by the Moroccan Ministry of Higher Education and Scientific Research MESRSFC and  CNRST: Projet PPR/2015/6. CHC was supported by the Ministry of Science and Technology of Taiwan,  under grant MOST-106-2112-M-006-010-MY2 (CHC). J.K. Parry was supported by the CAS PIFI program with Grant No. 2016PM020.

\appendix
%
\section{Loop functions}
We collect in this appendix the various functions appearing in the processes computed in the text with two variables obtained from the penguin and box diagrams.
\begin{eqnarray}
 f_{1}(x,y)  &=& \frac{x}{72}\bigg[\frac{7 y^2 - 5 y x - 8 x^2 }{(y-x)^3} + 
\frac{6 y x (3 x - 2 y)}{(y-x)^4} \log\left(\frac{y}{x}\right)\bigg]\\
f_{2}(x,y)  &=& \frac{x}{12}\bigg[\frac{3 y - 5 x }{(y-x)^2} + 
\frac{2 y (3 x - 2 y)}{(x-y)^3} \log\left(\frac{x}{y}\right)\bigg]\\
f_{3}(x,y)  &=& \frac{x}{108}\bigg[\frac{38 y^2 - 79 x y + 47 x^2 }{(y-x)^3} -
\frac{6 (4 y^3 - 6 y^2 x + 3 x^3)}{(y-x)^4} \log\left(\frac{y}{x}\right)\bigg]\\
f_{4}(x,y)  &=& \frac{x}{108}\bigg[\frac{-37 y^2 + 8 x y + 53 x^2 }{(y-x)^4} +
\frac{6 (2 y^3 + 6 y^2 x  - 9 y x^2 - 3x^3)}{(y-x)^5} \log\left(\frac{y}{x}\right)\bigg]\\
f_{5}(x,y)  &=& \frac{x }{8(y-x)}\bigg[\frac{-1}{(y-1)} +
\frac{y (1-y)\log(x)}{(y-x)(x-1)(y-1)}  - \frac{y (x+1 - 2 y)\log(y)}{(y-x)(y-1)^2} \bigg]\\
f_{6}(x,y)  &=& \frac{x }{16}  \bigg[\frac{-1}{(y-x)} + \frac{x}{(y-x)^2} \log\left(\frac{y}{x}\right)\bigg]\\
f_{7}(x,y)  &=& \frac{x }{8} \bigg[\frac{(x+2)\log(x)}{(y-x)(x-1)} - \frac{(y+2)\log(y)}{(y-x)(y-1)}\bigg]\\
f_{8}(x,y)  &=& \frac{x^2}{8}\bigg[\frac{1}{(y-x)} - \frac{y}{(y-x)^2} \log\left(\frac{y}{x}\right)\bigg]\\
f_{9}(x,y)  &=& \frac{x}{16} \bigg[\frac{y+x}{(y-x)^2} - \frac{2 x y}{(y-x)^3} \log\left(\frac{y}{x}\right)\bigg]\\
f_{10}(x,y)  &=& \frac{x}{16} \bigg[\frac{-1}{(y-x)} + \frac{x}{(y-x)^2} \log\left(\frac{y}{x}\right)\bigg]\\
f_{11}(x,y)  &=& \frac{x}{8} \bigg[\frac{(x-2)\log(x)}{(y-x)(x-1)} - \frac{(y-2)\log(y)}{(y-x)(y-1)} \bigg]\\\nonumber
\end{eqnarray}
\begin{eqnarray}
g_1(x,y) &=& \frac{x }{16}\bigg[ \frac{x-3y}{(y-x)^2} + \frac{2y^2}{(y-x)^3}\log\bigg(\frac{y}{x}\bigg)\bigg]\\
g_2(x,y) &=& \frac{x }{216}\bigg[ \frac{38y^2 + 54 y^2 x - 79yx - 108yx^2 + 47x^2 + 54x^3}{(y-x)^3} \\ &-& \frac{6(4y^3 + 9y^3 x - 6y^2 x - 18 y^2 x^2 + 9 y x^3 + 3 x^3)}{(y-x)^4}\log\bigg(\frac{y}{x}\bigg)\bigg]\\
g_3(x,y) &=& \frac{3 x }{432}\bigg[ \frac{2y^2 + 36y^2 x - 7yx - 72yx^2 + 11x^2 + 36x^3}{(y-x)^3} \\ &-& \frac{6 x (6 y^3 - 12y^2 x + 6y x^2 + x^2)}{(y-x)^4}\log\bigg(\frac{y}{x}\bigg)\bigg]\\
g_4(x,y) &=& \frac{x }{8(y-x)}\bigg[ \frac{x}{x-1}\log(x) - \frac{y}{y-1}\log(y)\bigg]\\
g_5(x,y) &=& \frac{x }{8(y-x)}\bigg[ 1 - \frac{y-x^2}{(x-1)(y-x)}\log(x) - \frac{y(x-1)}{(y-1)(y-x)}\log(y)\bigg]\\
g_6(x,y) &=& \frac{x}{8(y-x)}\log\left( \frac{x}{y}\right)\\\nonumber
\end{eqnarray}

\clearpage

\bibliographystyle{unsrt}

\begin{thebibliography}{100}


\expandafter\ifx\csname natexlab\endcsname\relax\def\natexlab#1{#1}\fi
\expandafter\ifx\csname bibnamefont\endcsname\relax
  \def\bibnamefont#1{#1}\fi
\expandafter\ifx\csname bibfnamefont\endcsname\relax
  \def\bibfnamefont#1{#1}\fi
\expandafter\ifx\csname citenamefont\endcsname\relax
  \def\citenamefont#1{#1}\fi
\expandafter\ifx\csname url\endcsname\relax
  \def\url#1{\texttt{#1}}\fi
\expandafter\ifx\csname urlprefix\endcsname\relax\def\urlprefix{URL }\fi
\providecommand{\bibinfo}[2]{#2}
\providecommand{\eprint}[2][]{\url{#2}}



\bibitem[{\citenamefont{Aad et~al.}(2012)}]{Aad:2012tfa}
\bibinfo{author}{\bibfnamefont{G.}~\bibnamefont{Aad}} \bibnamefont{et~al.}
  (\bibinfo{collaboration}{ATLAS Collaboration}), \bibinfo{journal}{Phys.Lett.}
  \textbf{\bibinfo{volume}{B716}}, \bibinfo{pages}{1} (\bibinfo{year}{2012}),
  \eprint{1207.7214}.


\bibitem[{\citenamefont{Chatrchyan et~al.}(2012)}]{Chatrchyan:2012ufa}
\bibinfo{author}{\bibfnamefont{S.}~\bibnamefont{Chatrchyan}}
  \bibnamefont{et~al.} (\bibinfo{collaboration}{CMS Collaboration}),
  \bibinfo{journal}{Phys.Lett.} \textbf{\bibinfo{volume}{B716}},
  \bibinfo{pages}{30} (\bibinfo{year}{2012}), \eprint{1207.7235}.



\bibitem[{\citenamefont{Egede et~al.}(2008)\citenamefont{Egede, Hurth, Matias,
  Ramon, and Reece}}]{Egede:2008uy}
\bibinfo{author}{\bibfnamefont{U.}~\bibnamefont{Egede}},
  \bibinfo{author}{\bibfnamefont{T.}~\bibnamefont{Hurth}},
  \bibinfo{author}{\bibfnamefont{J.}~\bibnamefont{Matias}},
  \bibinfo{author}{\bibfnamefont{M.}~\bibnamefont{Ramon}}, \bibnamefont{and}
  \bibinfo{author}{\bibfnamefont{W.}~\bibnamefont{Reece}},
  \bibinfo{journal}{JHEP} \textbf{\bibinfo{volume}{0811}}, \bibinfo{pages}{032}
  (\bibinfo{year}{2008}), \eprint{0807.2589}.

  
 
\bibitem[{\citenamefont{Aaij et~al.}(2013{\natexlab{a}})}]{Aaij:2013qta}
\bibinfo{author}{\bibfnamefont{R.}~\bibnamefont{Aaij}} \bibnamefont{et~al.}
(\bibinfo{collaboration}{LHCb collaboration}),
\bibinfo{journal}{Phys.Rev.Lett.} \textbf{\bibinfo{volume}{111}},
\bibinfo{pages}{191801} (\bibinfo{year}{2013}{\natexlab{a}}),
\eprint{1308.1707}. 
  

\bibitem[{\citenamefont{Collaboration}(2015)}]{LHCb:2015dla}
\bibinfo{author}{\bibfnamefont{T.~L.} \bibnamefont{Collaboration}}
  (\bibinfo{collaboration}{LHCb}) (\bibinfo{year}{2015}).
  
  
\bibitem{Khachatryan:2015isa}
  V.~Khachatryan {\it et al.} [CMS Collaboration],
  Phys.\ Lett.\ B {\bf 753} (2016) 424, 
    [\href{https://arxiv.org/abs/1507.08126}{{\tt hep-ph/1507.08126}}]
  
 

\bibitem{Wehle:2016yoi}
  S.~Wehle {\it et al.} [Belle Collaboration],
  Phys.\ Rev.\ Lett.\  {\bf 118} (2017) no.11,  111801,
  [\href{https://arxiv.org/abs/1612.05014}{{\tt hep-ph/1612.05014}}]
  
  


\bibitem[{\citenamefont{Descotes-Genon
  et~al.}(2013{\natexlab{a}})\citenamefont{Descotes-Genon, Hurth, Matias, and
  Virto}}]{Descotes-Genon:2013vna}
\bibinfo{author}{\bibfnamefont{S.}~\bibnamefont{Descotes-Genon}},
  \bibinfo{author}{\bibfnamefont{T.}~\bibnamefont{Hurth}},
  \bibinfo{author}{\bibfnamefont{J.}~\bibnamefont{Matias}}, \bibnamefont{and}
  \bibinfo{author}{\bibfnamefont{J.}~\bibnamefont{Virto}},
  \bibinfo{journal}{JHEP} \textbf{\bibinfo{volume}{1305}}, \bibinfo{pages}{137}
  (\bibinfo{year}{2013}{\natexlab{a}}), \eprint{1303.5794}.
  

\bibitem[{\citenamefont{Descotes-Genon
  et~al.}(2014)\citenamefont{Descotes-Genon, Hofer, Matias, and
  Virto}}]{Descotes-Genon:2014uoa}
\bibinfo{author}{\bibfnamefont{S.}~\bibnamefont{Descotes-Genon}},
  \bibinfo{author}{\bibfnamefont{L.}~\bibnamefont{Hofer}},
  \bibinfo{author}{\bibfnamefont{J.}~\bibnamefont{Matias}}, \bibnamefont{and}
  \bibinfo{author}{\bibfnamefont{J.}~\bibnamefont{Virto}},
  \bibinfo{journal}{JHEP} \textbf{\bibinfo{volume}{1412}}, \bibinfo{pages}{125}
  (\bibinfo{year}{2014}), \eprint{1407.8526}.
  
    

\bibitem{Altmannshofer:2014rta}
  W.~Altmannshofer and D.~M.~Straub,
  Eur.\ Phys.\ J.\ C {\bf 75} (2015) no.8,  382,
   [\href{https://arxiv.org/abs/1411.3161}{{\tt hep-ph/1411.3161}}]
  
     

\bibitem{Jager:2014rwa}
  S.~Jäger and J.~Martin Camalich,
  Phys.\ Rev.\ D {\bf 93} (2016) no.1,  014028,
  [\href{https://arxiv.org/abs/1412.3183}{{\tt hep-ph/1412.3183}}]
  


\bibitem[{\citenamefont{Aaij et~al.}(2013{\natexlab{b}})}]{Aaij:2013aln}
\bibinfo{author}{\bibfnamefont{R.}~\bibnamefont{Aaij}} \bibnamefont{et~al.}
  (\bibinfo{collaboration}{LHCb}), \bibinfo{journal}{JHEP}
  \textbf{\bibinfo{volume}{1307}}, \bibinfo{pages}{084}
  (\bibinfo{year}{2013}{\natexlab{b}}), \eprint{1305.2168}.
  
 
 
\bibitem[{\citenamefont{Horgan et~al.}(2014{\natexlab{a}})\citenamefont{Horgan,
Liu, Meinel, and Wingate}}]{Horgan:2013pva}
\bibinfo{author}{\bibfnamefont{R.~R.} \bibnamefont{Horgan}},
  \bibinfo{author}{\bibfnamefont{Z.}~\bibnamefont{Liu}},
  \bibinfo{author}{\bibfnamefont{S.}~\bibnamefont{Meinel}}, \bibnamefont{and}
  \bibinfo{author}{\bibfnamefont{M.}~\bibnamefont{Wingate}},
  \bibinfo{journal}{Phys.Rev.Lett.} \textbf{\bibinfo{volume}{112}},
  \bibinfo{pages}{212003} (\bibinfo{year}{2014}{\natexlab{a}}),
\eprint{1310.3887}.
  

\bibitem[{\citenamefont{Horgan et~al.}(2015)\citenamefont{Horgan, Liu, Meinel,
and Wingate}}]{Horgan:2015vla}
\bibinfo{author}{\bibfnamefont{R.}~\bibnamefont{Horgan}},
  \bibinfo{author}{\bibfnamefont{Z.}~\bibnamefont{Liu}},
  \bibinfo{author}{\bibfnamefont{S.}~\bibnamefont{Meinel}}, \bibnamefont{and}
  \bibinfo{author}{\bibfnamefont{M.}~\bibnamefont{Wingate}}
  (\bibinfo{year}{2015}), \eprint{1501.00367}.       
    

\bibitem[{\citenamefont{Bharucha et~al.}(2015)\citenamefont{Bharucha, Straub,
  and Zwicky}}]{Straub:2015ica}
\bibinfo{author}{\bibfnamefont{A.}~\bibnamefont{Bharucha}},
  \bibinfo{author}{\bibfnamefont{D.~M.} \bibnamefont{Straub}},
  \bibnamefont{and} \bibinfo{author}{\bibfnamefont{R.}~\bibnamefont{Zwicky}}
  (\bibinfo{year}{2015}), \eprint{1503.05534}.
  

\bibitem[{\citenamefont{Aaij et~al.}(2014)}]{Aaij:2014ora}
\bibinfo{author}{\bibfnamefont{R.}~\bibnamefont{Aaij}} \bibnamefont{et~al.}
  (\bibinfo{collaboration}{LHCb collaboration}),
  \bibinfo{journal}{Phys.Rev.Lett.} \textbf{\bibinfo{volume}{113}},
  \bibinfo{pages}{151601} (\bibinfo{year}{2014}), \eprint{1406.6482}.     
 

\bibitem[{\citenamefont{Bobeth et~al.}(2007)\citenamefont{Bobeth, Hiller, and
  Piranishvili}}]{Bobeth:2007dw}
\bibinfo{author}{\bibfnamefont{C.}~\bibnamefont{Bobeth}},
  \bibinfo{author}{\bibfnamefont{G.}~\bibnamefont{Hiller}}, \bibnamefont{and}
  \bibinfo{author}{\bibfnamefont{G.}~\bibnamefont{Piranishvili}},
  \bibinfo{journal}{JHEP} \textbf{\bibinfo{volume}{0712}}, \bibinfo{pages}{040}
  (\bibinfo{year}{2007}), \eprint{0709.4174}.
  


\bibitem{Altmannshofer:2017fio}
  W.~Altmannshofer, C.~Niehoff, P.~Stangl and D.~M.~Straub,
  Eur.\ Phys.\ J.\ C {\bf 77} (2017) no.6,  377
  doi:10.1140/epjc/s10052-017-4952-0
  [arXiv:1703.09189 [hep-ph]].
  
     

\bibitem[{\citenamefont{Altmannshofer and
  Straub}(2015)}]{Altmannshofer:2015sma}
\bibinfo{author}{\bibfnamefont{W.}~\bibnamefont{Altmannshofer}}
  \bibnamefont{and} \bibinfo{author}{\bibfnamefont{D.~M.} \bibnamefont{Straub}}
  (\bibinfo{year}{2015}), \eprint{1503.06199}. 
 



\bibitem{Hurth:2014vma} 
  T.~Hurth, F.~Mahmoudi and S.~Neshatpour,
  JHEP {\bf 1412}, 053 (2014)
  doi:10.1007/JHEP12(2014)053
  [arXiv:1410.4545 [hep-ph]].
 

\bibitem{Hiller:2014yaa} 
  G.~Hiller and M.~Schmaltz,
  Phys.\ Rev.\ D {\bf 90}, 054014 (2014)
  doi:10.1103/PhysRevD.90.054014
  [arXiv:1408.1627 [hep-ph]].
  
  

\bibitem[{\citenamefont{Descotes-Genon
  et~al.}(2013{\natexlab{b}})\citenamefont{Descotes-Genon, Matias, and
  Virto}}]{Descotes-Genon:2013wba}
\bibinfo{author}{\bibfnamefont{S.}~\bibnamefont{Descotes-Genon}},
  \bibinfo{author}{\bibfnamefont{J.}~\bibnamefont{Matias}}, \bibnamefont{and}
  \bibinfo{author}{\bibfnamefont{J.}~\bibnamefont{Virto}},
  \bibinfo{journal}{Phys.Rev.} \textbf{\bibinfo{volume}{D88}},
  \bibinfo{pages}{074002} (\bibinfo{year}{2013}{\natexlab{b}}),
  \eprint{1307.5683}. 
  



\bibitem[{\citenamefont{Gauld et~al.}(2014{\natexlab{a}})\citenamefont{Gauld,
  Goertz, and Haisch}}]{Gauld:2013qba}
\bibinfo{author}{\bibfnamefont{R.}~\bibnamefont{Gauld}},
  \bibinfo{author}{\bibfnamefont{F.}~\bibnamefont{Goertz}}, \bibnamefont{and}
  \bibinfo{author}{\bibfnamefont{U.}~\bibnamefont{Haisch}},
  \bibinfo{journal}{Phys.Rev.} \textbf{\bibinfo{volume}{D89}},
  \bibinfo{pages}{015005} (\bibinfo{year}{2014}{\natexlab{a}}),
  \eprint{1308.1959}.
  

\bibitem[{\citenamefont{Buras and Girrbach}(2013)}]{Buras:2013qja}
\bibinfo{author}{\bibfnamefont{A.~J.} \bibnamefont{Buras}} \bibnamefont{and}
  \bibinfo{author}{\bibfnamefont{J.}~\bibnamefont{Girrbach}},
  \bibinfo{journal}{JHEP} \textbf{\bibinfo{volume}{1312}}, \bibinfo{pages}{009}
  (\bibinfo{year}{2013}), \eprint{1309.2466}.
  

\bibitem[{\citenamefont{Gauld et~al.}(2014{\natexlab{b}})\citenamefont{Gauld,
  Goertz, and Haisch}}]{Gauld:2013qja}
\bibinfo{author}{\bibfnamefont{R.}~\bibnamefont{Gauld}},
  \bibinfo{author}{\bibfnamefont{F.}~\bibnamefont{Goertz}}, \bibnamefont{and}
  \bibinfo{author}{\bibfnamefont{U.}~\bibnamefont{Haisch}},
  \bibinfo{journal}{JHEP} \textbf{\bibinfo{volume}{1401}}, \bibinfo{pages}{069}
  (\bibinfo{year}{2014}{\natexlab{b}}), \eprint{1310.1082}.
  

\bibitem[{\citenamefont{Buras et~al.}(2014)\citenamefont{Buras, De~Fazio, and
  Girrbach}}]{Buras:2013dea}
\bibinfo{author}{\bibfnamefont{A.~J.} \bibnamefont{Buras}},
  \bibinfo{author}{\bibfnamefont{F.}~\bibnamefont{De~Fazio}}, \bibnamefont{and}
  \bibinfo{author}{\bibfnamefont{J.}~\bibnamefont{Girrbach}},
  \bibinfo{journal}{JHEP} \textbf{\bibinfo{volume}{1402}}, \bibinfo{pages}{112}
  (\bibinfo{year}{2014}), \eprint{1311.6729}.
  

\bibitem[{\citenamefont{Altmannshofer et~al.}(2014)\citenamefont{Altmannshofer,
  Gori, Pospelov, and Yavin}}]{Altmannshofer:2014cfa}
\bibinfo{author}{\bibfnamefont{W.}~\bibnamefont{Altmannshofer}},
  \bibinfo{author}{\bibfnamefont{S.}~\bibnamefont{Gori}},
  \bibinfo{author}{\bibfnamefont{M.}~\bibnamefont{Pospelov}}, \bibnamefont{and}
  \bibinfo{author}{\bibfnamefont{I.}~\bibnamefont{Yavin}},
  \bibinfo{journal}{Phys.Rev.} \textbf{\bibinfo{volume}{D89}},
  \bibinfo{pages}{095033} (\bibinfo{year}{2014}), \eprint{1403.1269}.
  
 
\bibitem{Crivellin:2015mga} 
  A.~Crivellin, G.~D'Ambrosio and J.~Heeck,
  Phys.\ Rev.\ Lett.\  {\bf 114}, 151801 (2015)
  doi:10.1103/PhysRevLett.114.151801
  [arXiv:1501.00993 [hep-ph]].
  
  

\bibitem[{\citenamefont{Crivellin
  et~al.}(2015{\natexlab{b}})\citenamefont{Crivellin, D'Ambrosio, and
  Heeck}}]{Crivellin:2015lwa}
\bibinfo{author}{\bibfnamefont{A.}~\bibnamefont{Crivellin}},
  \bibinfo{author}{\bibfnamefont{G.}~\bibnamefont{D'Ambrosio}},
  \bibnamefont{and} \bibinfo{author}{\bibfnamefont{J.}~\bibnamefont{Heeck}}
  (\bibinfo{year}{2015}{\natexlab{b}}), \eprint{1503.03477}.
  

\bibitem[{\citenamefont{Niehoff et~al.}(2015)\citenamefont{Niehoff, Stangl, and
  Straub}}]{Niehoff:2015bfa}
\bibinfo{author}{\bibfnamefont{C.}~\bibnamefont{Niehoff}},
  \bibinfo{author}{\bibfnamefont{P.}~\bibnamefont{Stangl}}, \bibnamefont{and}
  \bibinfo{author}{\bibfnamefont{D.~M.} \bibnamefont{Straub}}
  (\bibinfo{year}{2015}), \eprint{1503.03865}.
  

\bibitem[{\citenamefont{Sierra et~al.}(2015)\citenamefont{Sierra, Staub, and
  Vicente}}]{Sierra:2015fma}
\bibinfo{author}{\bibfnamefont{D.~A.} \bibnamefont{Sierra}},
  \bibinfo{author}{\bibfnamefont{F.}~\bibnamefont{Staub}}, \bibnamefont{and}
  \bibinfo{author}{\bibfnamefont{A.}~\bibnamefont{Vicente}}
  (\bibinfo{year}{2015}), \eprint{1503.06077}.
  

\bibitem[{\citenamefont{Celis et~al.}(2015)\citenamefont{Celis, Fuentes-Martin,
  Jung, and Serodio}}]{Celis:2015ara}
\bibinfo{author}{\bibfnamefont{A.}~\bibnamefont{Celis}},
  \bibinfo{author}{\bibfnamefont{J.}~\bibnamefont{Fuentes-Martin}},
  \bibinfo{author}{\bibfnamefont{M.}~\bibnamefont{Jung}}, \bibnamefont{and}
  \bibinfo{author}{\bibfnamefont{H.}~\bibnamefont{Serodio}}
  (\bibinfo{year}{2015}), \eprint{1505.03079}.
  

\bibitem[{\citenamefont{Becirevic et~al.}(2015)\citenamefont{Becirevic, Fajfer,
  and Kosnik}}]{Becirevic:2015asa}
\bibinfo{author}{\bibfnamefont{D.}~\bibnamefont{Becirevic}},
  \bibinfo{author}{\bibfnamefont{S.}~\bibnamefont{Fajfer}}, \bibnamefont{and}
  \bibinfo{author}{\bibfnamefont{N.}~\bibnamefont{Kosnik}}
  (\bibinfo{year}{2015}), \eprint{1503.09024}.
  

\bibitem[{\citenamefont{Varzielas and Hiller}(2015)}]{Varzielas:2015iva}
\bibinfo{author}{\bibfnamefont{I.~d.~M.} \bibnamefont{Varzielas}}
  \bibnamefont{and} \bibinfo{author}{\bibfnamefont{G.}~\bibnamefont{Hiller}}
  (\bibinfo{year}{2015}), \eprint{1503.01084}.
  

\bibitem{Glashow:2014iga} 
  S.~L.~Glashow, D.~Guadagnoli and K.~Lane,
  Phys.\ Rev.\ Lett.\  {\bf 114}, 091801 (2015)
  doi:10.1103/PhysRevLett.114.091801
  [arXiv:1411.0565 [hep-ph]].
  
              
\bibitem{Boucenna:2015raa} 
  S.~M.~Boucenna, J.~W.~F.~Valle and A.~Vicente,
  Phys.\ Lett.\ B {\bf 750}, 367 (2015)
  doi:10.1016/j.physletb.2015.09.040
  [arXiv:1503.07099 [hep-ph]].
  

\bibitem{Bobeth:2001sq} 
  C.~Bobeth, T.~Ewerth, F.~Kruger and J.~Urban,
  Phys.\ Rev.\ D {\bf 64}, 074014 (2001)
  doi:10.1103/PhysRevD.64.074014
  [hep-ph/0104284].
  
  

          
      
\bibitem{Jung:2012vu}
  M.~Jung, X.~Q.~Li and A.~Pich,
  JHEP {\bf 1210} (2012) 063
  doi:10.1007/JHEP10(2012)063
  [arXiv:1208.1251 [hep-ph]].
  
  

\bibitem{Hu:2016gpe}
  Q.~Y.~Hu, X.~Q.~Li and Y.~D.~Yang,
  Eur.\ Phys.\ J.\ C {\bf 77} (2017) no.3,  190
  doi:10.1140/epjc/s10052-017-4748-2
  [arXiv:1612.08867 [hep-ph]].
    

\bibitem{Arnan:2017lxi} 
  P.~Arnan, D.~Bečirević, F.~Mescia and O.~Sumensari,
  arXiv:1703.03426 [hep-ph].


\bibitem{Branco:2011iw} 
  G.~C.~Branco, P.~M.~Ferreira, L.~Lavoura, M.~N.~Rebelo, M.~Sher and J.~P.~Silva,
  Phys.\ Rept.\  {\bf 516}, 1 (2012)
  doi:10.1016/j.physrep.2012.02.002
  [arXiv:1106.0034 [hep-ph]].


\bibitem{hunter} 
  J.~F.~Gunion, H.~E.~Haber, G.~L.~Kane and S.~Dawson,
  Front.\ Phys.\  {\bf 80}, 1 (2000).




\bibitem{Cheng:1987rs} 
  T.~P.~Cheng and M.~Sher,
  Phys.\ Rev.\ D {\bf 35}, 3484 (1987).


\bibitem{Alonso:2016oyd} 
  R.~Alonso, B.~Grinstein and J.~Martin Camalich,
  Phys.\ Rev.\ Lett.\  {\bf 118}, no. 8, 081802 (2017), 
  [\href{https://arxiv.org/abs/1611.06676}{{\tt hep-ph/1611.06676}}]


\bibitem{PDG} C. Patrignani {\it et al.} (Particle Data Group), Chin.\ Phys.\ C {\bf 40}, 100001 (2016).




\bibitem{Lenz:2010gu} 
  A.~Lenz {\it et al.},
  Phys.\ Rev.\ D {\bf 83}, 036004 (2011)
[\href{https://arxiv.org/abs/1008.1593}{{\tt hep-ph/1008.1593}}]


\bibitem{Colquhoun:2015oha} 
  B.~Colquhoun {\it et al.} [HPQCD Collaboration],
  Phys.\ Rev.\ D {\bf 91}, no. 11, 114509 (2015)
  [\href{https://arxiv.org/abs/1503.05762}{{\tt hep-lat/1503.05762}}]


\bibitem{Becirevic:2001jj} 
  D.~Becirevic {\it et al.},
  Nucl.\ Phys.\ B {\bf 634}, 105 (2002),
  [\href{https://arxiv.org/abs/hep-ph/0112303}{{\tt hep-ph/0112303}}]


\bibitem{Buchalla:1995vs} 
  G.~Buchalla, A.~J.~Buras and M.~E.~Lautenbacher,
  Rev.\ Mod.\ Phys.\  {\bf 68}, 1125 (1996)
   [\href{https://arxiv.org/abs/hep-ph/9512380}{{\tt hep-ph/9512380}}]


\bibitem{Inami:1980fz} 
  T.~Inami and C.~S.~Lim,
  Prog.\ Theor.\ Phys.\  {\bf 65}, 297 (1981)
  Erratum: [Prog.\ Theor.\ Phys.\  {\bf 65}, 1772 (1981)].


\bibitem{Urban:1997gw} 
  J.~Urban, F.~Krauss, U.~Jentschura and G.~Soff,
  Nucl.\ Phys.\ B {\bf 523}, 40 (1998),
[\href{https://arxiv.org/abs/hep-ph/9710245}{{\tt hep-ph/9710245}}]


\bibitem{Becirevic:2001xt} 
  D.~Becirevic, V.~Gimenez, G.~Martinelli, M.~Papinutto and J.~Reyes,
  JHEP {\bf 0204}, 025 (2002),
  [\href{https://arxiv.org/abs/hep-lat/0110091}{{\tt hep-lat/0110091}}]


\bibitem{Becirevic:2001yv} 
  D.~Becirevic, V.~Gimenez, G.~Martinelli, M.~Papinutto and J.~Reyes,
  Nucl.\ Phys.\ Proc.\ Suppl.\  {\bf 106}, 385 (2002), 
   [\href{https://arxiv.org/abs/hep-lat/0110117}{{\tt hep-lat/0110117}}] 



\bibitem{Buras:1990fn} 
  A.~J.~Buras, M.~Jamin and P.~H.~Weisz,
  Nucl.\ Phys.\ B {\bf 347}, 491 (1990).
  
  
\bibitem{Ciuchini:1997bw} 
  M.~Ciuchini, E.~Franco, V.~Lubicz, G.~Martinelli, I.~Scimemi and L.~Silvestrini,
  Nucl.\ Phys.\ B {\bf 523}, 501 (1998), 
     [\href{https://arxiv.org/abs/hep-ph/9711402}{{\tt hep-ph/9711402}}] 


\bibitem{Buras:2000if} 
  A.~J.~Buras, M.~Misiak and J.~Urban,
  Nucl.\ Phys.\ B {\bf 586}, 397 (2000), 
 [\href{https://arxiv.org/abs/hep-ph/0005183}{{\tt hep-ph/0005183}}] 


 

\bibitem{Lees:2013nxa} 
  J.~P.~Lees {\it et al.} [BaBar Collaboration],
  Phys.\ Rev.\ Lett.\  {\bf 112}, 211802 (2014)
  doi:10.1103/PhysRevLett.112.211802
  [arXiv:1312.5364 [hep-ex]].

\bibitem{Patrignani:2016xqp}
  C.~Patrignani {\it et al.} [Particle Data Group],
  Chin.\ Phys.\ C {\bf 40} (2016) no.10,  100001.
  doi:10.1088/1674-1137/40/10/100001

\bibitem{Amhis:2016xyh} 
  Y.~Amhis {\it et al.},
  [\href{https://arxiv.org/abs/1612.07233}{{\tt hep-ph/1612.07233}}]


\bibitem{Czakon:2015exa} 
  M.~Czakon, P.~Fiedler, T.~Huber, M.~Misiak, T.~Schutzmeier and M.~Steinhauser,
  JHEP {\bf 1504}, 168 (2015), 
 [\href{https://arxiv.org/abs/1503.01791}{{\tt hep-ph/1503.01791}}]



\bibitem{Misiak:2015xwa} 
  M.~Misiak {\it et al.},
  Phys.\ Rev.\ Lett.\  {\bf 114}, no. 22, 221801 (2015), 
  [\href{https://arxiv.org/abs/1503.01789}{{\tt hep-ph/1503.01789}}]


\bibitem{Borzumati:1998tg} 
  F.~Borzumati and C.~Greub,
  Phys.\ Rev.\ D {\bf 58}, 074004 (1998),
[\href{https://arxiv.org/abs/hep-ph/9802391}{{\tt hep-ph/9802391}}]


\bibitem{Ciuchini:1997xe} 
  M.~Ciuchini, G.~Degrassi, P.~Gambino and G.~F.~Giudice,
  Nucl.\ Phys.\ B {\bf 527}, 21 (1998), 
 [\href{https://arxiv.org/abs/hep-ph/9710335}{{\tt hep-ph/9710335}}]



\bibitem{Borzumati:1998nx} 
  F.~Borzumati and C.~Greub,
  Phys.\ Rev.\ D {\bf 59}, 057501 (1999),
[\href{https://arxiv.org/abs/hep-ph/9809438}{{\tt hep-ph/9809438}}]


\bibitem{Hermann:2012fc} 
  T.~Hermann, M.~Misiak and M.~Steinhauser,
  JHEP {\bf 1211}, 036 (2012), 
[\href{https://arxiv.org/abs/1208.2788}{{\tt hep-ph/1208.2788}}]


\bibitem{Blanke:2011ry} 
  M.~Blanke, A.~J.~Buras, K.~Gemmler and T.~Heidsieck,
  JHEP {\bf 1203}, 024 (2012),
[\href{https://arxiv.org/abs/1111.5014}{{\tt hep-ph/1111.5014}}]


\bibitem{Misiak:2017bgg} 
  M.~Misiak and M.~Steinhauser,
  Eur.\ Phys.\ J.\ C {\bf 77}, no. 3, 201 (2017),
  [\href{https://arxiv.org/abs/1702.04571}{{\tt hep-ph/1702.04571}}]


\bibitem{Chetyrkin:1996vx} 
  K.~G.~Chetyrkin, M.~Misiak and M.~Munz,
  Phys.\ Lett.\ B {\bf 400}, 206 (1997),
  Erratum: [Phys.\ Lett.\ B {\bf 425}, 414 (1998)],
  [\href{https://arxiv.org/abs/hep-ph/9612313}{{\tt hep-ph/9612313}}]


\bibitem{Altmannshofer:2008dz} 
  W.~Altmannshofer, P.~Ball, A.~Bharucha, A.~J.~Buras, D.~M.~Straub and M.~Wick,
  JHEP {\bf 0901}, 019 (2009),
   [\href{https://arxiv.org/abs/0811.1214}{{\tt hep-ph/0811.1214}}]


\bibitem{DescotesGenon:2011yn} 
  S.~Descotes-Genon, D.~Ghosh, J.~Matias and M.~Ramon,
  JHEP {\bf 1106}, 099 (2011),
  [\href{https://arxiv.org/abs/1104.3342}{{\tt hep-ph/1104.3342}}]


\bibitem{Bobeth:2003at} 
  C.~Bobeth, P.~Gambino, M.~Gorbahn and U.~Haisch,
  JHEP {\bf 0404}, 071 (2004),
  [\href{https://arxiv.org/abs/hep-ph/0312090}{{\tt hep-ph/0312090}}]


\bibitem{Fleischer:2017ltw}
  R.~Fleischer, R.~Jaarsma and G.~Tetlalmatzi-Xolocotzi,
  JHEP {\bf 1705} (2017) 156
  doi:10.1007/JHEP05(2017)156
  [arXiv:1703.10160 [hep-ph]].
  R.~Fleischer, D.~G.~Espinosa, R.~Jaarsma and G.~Tetlalmatzi-Xolocotzi,
  arXiv:1709.04735 [hep-ph].

\bibitem{Li:2014jsa}
 X.~Q.~Li, J.~Lu and A.~Pich,
  JHEP {\bf 1406} (2014) 022
  doi:10.1007/JHEP06(2014)022
  [arXiv:1404.5865 [hep-ph]].
  X.~Q.~Li, J.~Lu and A.~Pich,
  Nucl.\ Part.\ Phys.\ Proc.\  {\bf 273-275} (2016) 1411
  doi:10.1016/j.nuclphysbps.2015.09.228
  [arXiv:1410.4775 [hep-ph]].

\bibitem{Huber:2005ig} 
  T.~Huber, E.~Lunghi, M.~Misiak and D.~Wyler,
  Nucl.\ Phys.\ B {\bf 740}, 105 (2006),
    [\href{https://arxiv.org/abs/hep-ph/0512066}{{\tt hep-ph/0512066}}]


\bibitem{Bobeth:1999mk} 
  C.~Bobeth, M.~Misiak and J.~Urban,
  Nucl.\ Phys.\ B {\bf 574}, 291 (2000),
  [\href{https://arxiv.org/abs/hep-ph/9910220}{{\tt hep-ph/9910220}}]
  


\bibitem{Aaij:2013iag} 
  R.~Aaij {\it et al.} [LHCb Collaboration],
  JHEP {\bf 1308}, 131 (2013),
   [\href{https://arxiv.org/abs/1304.6325}{{\tt hep-ph/1304.6325}}]



\bibitem{Aaij:2015oid}
  R.~Aaij {\it et al.} [LHCb Collaboration],
  JHEP {\bf 1602} (2016) 104,
  [\href{https://arxiv.org/abs/1512.04442}{{\tt hep-ph/1512.04442}}]



\bibitem{Chatrchyan:2013cda}
  S.~Chatrchyan {\it et al.} [CMS Collaboration],
  Phys.\ Lett.\ B {\bf 727} (2013) 77, 
   [\href{https://arxiv.org/abs/1308.3409}{{\tt hep-ph/1308.3409}}]



\bibitem{Aaltonen:2011ja}
  T.~Aaltonen {\it et al.} [CDF Collaboration],
  Phys.\ Rev.\ Lett.\  {\bf 108} (2012) 081807, 
  [\href{https://arxiv.org/abs/1108.0695}{{\tt hep-ph/1108.0695}}]


\bibitem{Wei:2009zv}
  J.-T.~Wei {\it et al.} [Belle Collaboration],
  Phys.\ Rev.\ Lett.\  {\bf 103} (2009) 171801,
  [\href{https://arxiv.org/abs/0904.0770}{{\tt hep-ph/0904.0770}}]


\bibitem{Aubert:2006vb}
  B.~Aubert {\it et al.} [BaBar Collaboration],
  Phys.\ Rev.\ D {\bf 73} (2006) 092001,
  [\href{https://arxiv.org/abs/hep-ex/0604007}{{\tt hep-ex/0604007}}]



\bibitem{Descotes-Genon:2015uva}
  S.~Descotes-Genon, L.~Hofer, J.~Matias and J.~Virto,
  JHEP {\bf 1606} (2016) 092
  [\href{https://arxiv.org/abs/1510.04239}{{\tt hep-ph/1510.04239}}]
  

\bibitem{Beaujean:2013soa}
  F.~Beaujean, C.~Bobeth and D.~van Dyk,
  Eur.\ Phys.\ J.\ C {\bf 74} (2014) 2897
   Erratum: [Eur.\ Phys.\ J.\ C {\bf 74} (2014) 3179]
  doi:10.1140/epjc/s10052-014-2897-0, 10.1140/epjc/s10052-014-3179-6
  [arXiv:1310.2478 [hep-ph]].


\bibitem{Fleischer:2017yox} 
  R.~Fleischer, D.~G.~Espinosa, R.~Jaarsma and G.~Tetlalmatzi-Xolocotzi,
  arXiv:1709.04735 [hep-ph].

\end{thebibliography}

\end{document}